\documentclass{article}

\usepackage[preprint, nonatbib]{neurips_2023}

\usepackage[utf8]{inputenc} 
\usepackage[T1]{fontenc}    
\usepackage[breaklinks=true]{hyperref}
\usepackage{url}            
\usepackage{booktabs}       
\usepackage{amsfonts}       
\usepackage{nicefrac}       
\usepackage{microtype}      
\usepackage{xcolor}         
\usepackage{amssymb}
\usepackage{amsmath}
\usepackage{algpseudocode}
\usepackage{algorithm}
\usepackage{algorithmicx}
\usepackage{graphicx}
\usepackage{caption}
\usepackage{subcaption}
\newcommand{\indep}{\rotatebox[origin=c]{90}{$\models$}}
\title{Information-Theoretic Grid Topology Reconstruction using Low-Precision Smart Meter Data}

\author{%
  Daniel T. Speckhard\thanks{Current address: The Institute of Physics at Humboldt-Universit\"at zu Berlin, Zum Gro\ss en Windkanal 2, Berlin, 12489, Germany and The NOMAD Laboratory at the Fritz Haber Institute of the Max Planck Society, Faradayweg 4-6, Berlin, 14195, Germany} \\
  Dept. of Electrical Engineering, Stanford University, USA \\
  Institut für Physik, Humboldt-Universit\"at zu Berlin, Germany \\
  \texttt{dts@stanford.edu} 
}

\begin{document}

\maketitle

\begin{abstract}
Accurate knowledge of power grid topology is a prerequisite for effective state estimation and grid stability. While data-driven methods for topology reconstruction exist, the minimum requirements for measurement quality, specifically regarding quantization, precision, and sampling frequency, remain under-explored. This study investigates the data fidelity required to reconstruct distribution grid topologies using voltage magnitude measurements. Adopting an information-theoretic approach, we utilize the Chow-Liu algorithm to generate maximum spanning trees based on mutual information. Rather than proposing a new reconstruction algorithm, our primary contribution is a comprehensive sensitivity analysis of the measurement data itself. We systematically evaluate the impact of data bit-depth, significant digit truncation, time-window length, and different mutual information estimators on reconstruction accuracy. We validate this approach using IEEE test cases (via MATPOWER) and time-series data from GridLAB-D. Our results demonstrate that grid topology can be successfully recovered even with highly quantized 8-bit data or millivolt-level precision. However, performance degrades significantly when downsampling intervals exceed 20 minutes or when data availability is limited to short durations. These findings establish an optimistic theoretical lower bound, suggesting that costly high-precision instrumentation may not be strictly necessary for structural inference under ideal conditions. This rigorous baseline provides a foundation for future evaluations of noisy real world smart meter data and hybrid approaches that incorporate existing engineering priors.
\end{abstract}

\section{Introduction}
\label{S:1}

The integration of distributed energy resources (DERs), such as solar and wind power plants, is driving a fundamental transformation in electrical grids worldwide. These variable generation sources necessitate frequent reconfiguration of grid topology, often mediated by vast arrays of digitally controlled switches. In this dynamic environment, accurate knowledge of the grid topology is critical for energy conservation, cost reduction, and load stability. Furthermore, topological data underpins the simulation and control algorithms required to optimize power transmission and distribution~\cite{liao2018urban}.

Despite this need, utility companies frequently rely on outdated or mislabeled electrical grid maps~\cite{park2020learning}. Inaccurate topology data severely impairs a utility's ability to observe and optimize power flow. For example, the strategic placement of a 500 kWh flow-battery to improve load stability relies heavily on precise topological knowledge. In many scenarios, utilities possess only a verified partial topology~\cite{lampe2013power} and must infer the connectivity of outlying sub-sections. Moreover, in most operational settings utilities do not lack topology information entirely but rather possess engineering models and GIS records that become uncertain after switching events or asset replacements~\cite{park2020learning, deka2023learning}. While hybrid approaches that combine these engineering priors with meter data are highly practical, our work focuses on the purely data driven problem. By isolating the data driven component we establish an independent verification layer and a theoretical baseline for the absolute minimum data requirements needed before fusing with uncertain engineering priors.

Previously, researchers in Ref.~\cite{liao2015distribution, liao2016urban} have defined the problem statement for the estimation of the graph structure of a power grid using smart-meter data. We can define a distribution grid as a graph with edges (branches) that connect different nodes (buses). To utilize the time series data collected by smart meters, for a $M$ bus system, we construct a graphical model $ G = (S,E)$ where $ S = \{1,2,...M\}$ is the set of vertices, and $ E = \{e_{ik},i,k \in S\} $ represents the set of the undirected edges meaning $ e_{ik} = e_{ki}$. In this model, which is common in the literature~\cite{poudel2023topology, zhu2024ggnet, zhao2025power} a node corresponds to a bus in the physical layer and is modeled as a random variable $V_{i}$. The edge that connects node $i$ and $k$ represents the statistical dependence measured between bus $i$ and $k$. If we assume for a noiseless system that measurements are taken in steady state and all voltages are sinusoidal signals at the same frequency, then the voltage measurement at bus $i$ is defined as $ v_{i}[t] = \lvert v_{i}[t] \rvert e^{(j\delta_{i}[t])}$, where $\lvert v_{i}[t] \rvert$ is the voltage magnitude of node $i$ at time $t$ and $\delta_{i}[t]$ is the voltage phase at node $i$ at time $t$. This work seeks to reconstruct the grid bus topology using voltage magnitude data and phase data when available. In other words, given a sequence of historical measurements $\lvert v_{i}(t)\rvert$ we seek to find graph structure of the power grid. In some datasets we will also have access to sensor current data (magnitude and phase, $\delta_{i}[t]$ and ${i}[t]$ respectively) as well. 

The literature presents several different approaches to topology estimation. Graph neural networks have recently become very popular in a wide variety of applied fields from materials science~\cite{bechtel2025band, speckhard2025big}, chemistry~\cite{fung2021benchmarking}, social sciences~\cite{fan2019graph} and have found success in power grid applications~\cite{liao2021review}. These neural network approaches are, however, generally quite data-intensive~\cite{wu2020comprehensive} and compute-intensive despite recent work to minimize the training cost~\cite{speckhard2025analysis}. Furthermore, their black-box nature obscures the fundamental physical data requirements necessary for structural inference~\cite{misyris2019grid, kordowich2023physics, eeckhout2024improved}.

Conversely, analytical methods offer highly efficient and interpretable alternatives. Recent tutorials and comparative studies highlight that pairwise distance metrics based on graph distances often outperform standard graphical model reconstructions~\cite{deka2017structure, deka2023learning}. While the broader impacts of data quality and analytics on the grid have been benchmarked in foundational Department of Energy reports~\cite{stewart2018integrated}, a critical gap remains in understanding the absolute minimum measurement fidelity, specifically extreme quantization, precision, and temporal resolution, required to accurately estimate a grid's topology.

Rather than proposing a new estimation algorithm, the primary contribution of this work is a rigorous, empirical sensitivity analysis of these data quality limits. To establish a transparent and interpretable baseline, we focus on an established information-theoretic method~\cite{weng2016distributed}: computing pairwise mutual information of sensor data and utilizing the Chow-Liu algorithm~\cite{chow1968approximating, koller2009probabilistic} to reconstruct the power grid structure. It is important to note that in linear systems, the true conditional dependence structure, the Markov blanket, extends to a 2-hop neighborhood~\cite{bolognani2013identification, anguluri2021grid}. Consequently, to utilize the Chow-Liu algorithm for learning the topology, we explicitly assume that the dependencies between nodes that are two or more hops away are bounded within a small limit ($\epsilon$). Under this $\epsilon$-bounded assumption, the algorithm is employed to learn an approximate topology. By treating the algorithm as a constant, we systematically isolate and test the breaking points of data quality, varying the length of data, bit-depth resolution, significant digit truncation, downsampling rate and mutual information approximation algorithm.

These experiments are important for operators to understand what length and precision of data are needed to store~\cite{hernandez2025monitoring, engel2025real}. We also test different entropy approximations which speed up the method's computation time of the mutual information~\cite{chen2021efficient}. We present the following contributions~\cite{speckhard2025grid}.

\begin{itemize}

\item{The method predicts the grid's graph structure 100\% correctly for smaller networks but finds more difficulty with a larger 200+ node network (90\% successful detection).}
\item{Some datasets show that downsampling smart meter data from measurements every minute to every hour, reduces our successful edge detection by 30\% while other datasets show no issues with hourly measurements.}
\item{The length of smart-meter data plays a significant role. We show a
greater than a 5\% increase in successful reconstruction when using one month of data compared to one week of data. The improvement trails off after six months of data are used.}
\item{Data quality is important. Experiments to reduce the number of bits with which to store voltage magnitude data show a greater than 10\% increase in successful estimation when using 8-bit data precision compared to 4-bit precision. Retaining millivolts of voltage magnitude data precision compared to 10’s of millivolts of precision can increase performance from 5\% to
15\%.}
\end{itemize}

\section{Methods}

\subsection{Theoretical estimation approach}
\label{S:2}

We seek to find the grid topology from smart meter time series data of bus voltage measurements. We model the buses as random variables. Assuming node 1 is the slack bus with a constant unity magnitude and zero phase angle, it is omitted from the joint probability formulation; thus, our nodal index begins at $i=2$. We can use a joint probability distribution to represent the inter-dependency among the remaining buses:

\begin{equation}
\label{eq:joint_probability}
P({\mathbf{V_{S}}}) = P(V_{2}, V_{3}, .... V_{M})
\end{equation}

This equation can be further simplified using the chain rule:
\begin{equation}
\label{eq:joint_probability_chain}
P({\mathbf{V_{S}}}) = P(V_{2})P(V_{3}|V_{2}) .... P(V_{M}|V_{2},V_{3},...V_{M-1})
\end{equation}

\noindent Voltage measurements usually have irregular probability distributions. Instead of using raw voltage measurements directly, we use the incremental change of the voltage measurement to reconstruct the distribution grid topology \cite{chen2016quickest}. For the remainder of this paper, we let the random variable $V_{i}$ denote this incremental change in voltage at bus $i$.

In linear systems, it can be clearly shown that the conditional dependence structure (the Markov blanket) formally extends to a 2-hop neighborhood~\cite{deka2023learning, bolognani2013identification, anguluri2021grid}. However, to facilitate our baseline information-theoretic reconstruction, we make the simplifying assumption that dependencies between nodes that are two or more hops away are weak and bounded within a small limit ($\epsilon$). Under this $\epsilon$-bounded assumption, the change in voltage at a bus is treated as approximately conditionally independent of the change in voltages at non-neighboring buses, given the change in voltages of neighboring buses~\cite{liao2016urban}. Since voltage and current changes are tightly coupled via the grid's sparse admittance matrix, we can analyze current as a proxy for nodal dependencies. Figure~\ref{fg:current_MI} in the Appendix shows the pairwise mutual information of the change in current phasors. The relatively small mutual information beyond the 1-hop neighborhood physically supports this $\epsilon$-approximation for voltage. 

In a distribution grid, the correlation between interconnected neighboring buses is higher than that between non-neighbor buses. Therefore, invoking this bounded assumption, the voltage at node $i$ is conditionally independent of non-neighboring nodes given its neighbors: $ V_{i} \indep V_{j \notin N(i)} | V_{k\in N(i)}$, where $N(i)$ contains the vertices of the neighbors of bus $i$, i.e., $ N(i) = \{k \in S|e_{ik} \in E\} $.

Since the distribution grid is modeled here as an undirected Markov Random Field, the exact joint probability cannot be strictly factorized by the product of full conditionals. Instead, we formulate the pseudo-likelihood objective, which is a standard analytical approximation for Markov networks~\cite{besag1974spatial, koller2009probabilistic}:

\begin{equation}
\label{eq:first_order}
P({\mathbf{V_{S}}}) \approx \prod\limits_{i=2}^M P_{i}(V_{i} | V_{\mathbf{N(i)}}) 
\end{equation}

The pseudo-likelihood approximation in Equation \ref{eq:first_order} mathematically simplifies the task of finding the connectivity of bus $i$ to finding the neighboring nodes of bus $i$.

Previous work demonstrated that a maximum spanning tree algorithm using mutual information as edge weights optimally approximates $P(V_{S})$ in Equation (\ref{eq:joint_probability})~\cite{weng2016distributed}. We restate the theorem from this work below without proof (Theorem 1).

\newtheorem{theorem}{Theorem}
\label{th:mutual_information}
\begin{theorem}
In a radial distribution power grid, the mutual information-based maximum spanning tree algorithm finds an optimal approximate learning of $P(V_{S})$ and its associated topology connection, provided that dependencies between nodes two or more hops away are within an $\epsilon$-limit bound, and current injections are approximated as independent.
\end{theorem}

With these theorems in mind, we employ the Chow-Liu algorithm~\cite{chow1968approximating} for our topology estimation problem. The algorithm finds the maximum weight spanning tree based on mutual information between nodes. The mutual information is defined as in Equation (\ref{eq:MI}) for continuous random variables.

\begin{equation}
\label{eq:MI}
I(X,Y) = \int_{x \in X}\int_{y \in Y}  \!P(x,y)\log\frac{P(x,y)}{P(x)P(y)} \, \mathrm{d}x \mathrm{d}y 
\end{equation}   

The algorithm uses the mutual information as a weighting factor in the dependence tree it estimates. The mutual information was shown by Chow and Liu to optimally estimate the low order product approximation of a joint probability distribution. For our purposes the algorithm approximates Equation (\ref{eq:joint_probability}).

The algorithm proceeds in the following steps. 

\begin{algorithm}
\caption{Tree Structure Topology Reconstruction}
\label{alg:tree}
\begin{algorithmic}[1]
\Require : $v_{i}[t]$ for $i$ = 2, ..., $M$, $t$=1, ..., $T$
\For{$i$,$j =2$, ..., $M$} 

\State Compute mutual information $I(V_{i}, V_{j})$ based on $v_{i}[t]$. 
\EndFor
\State Sort all possible bus pairs $(i,j)$ into non-increasing order by $I(V_{i}, V_{j})$. Let $\tilde{\mathcal{E}}$ denote the sorted set.
\State Let $\hat{\mathcal{E}}$ be the set of nodal pairs comprising the maximum weight spanning tree. Set $\hat{\mathcal{E}} = \emptyset$
\For{$(i,j) \in \tilde{\mathcal{E}}$ } 
\If{cycle is detected in $\hat{\mathcal{E}} \cup (i,j)$} 

\State \textbf{Continue}
\Else 
\State $\hat{\mathcal{E}} \leftarrow \hat{\mathcal{E}} \cup (i,j)$
\EndIf
\If{ $\lvert \hat{\mathcal{E}} \rvert == M -2$}
\State \textbf{break}
\EndIf
\EndFor
\State \textbf{return $\hat{\mathcal{E}}$}
\end{algorithmic}
\end{algorithm}

\begin{figure}
\centering\includegraphics[width=1.0\linewidth]{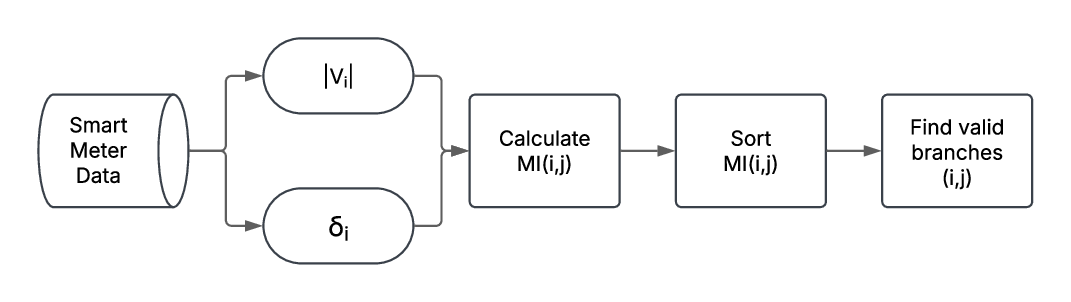}
\caption{The program flow of the Chow-Liu Algorithm applied to electrical grid topology estimation. When only voltage magnitude data is available, only the top half of the parallel split in the workflow is executed.}

\label{fg:algorithm_flow}
\end{figure}

We can visualize the algorithm steps in Figure \ref{fg:algorithm_flow} as a data analysis workflow (DAW)~\cite{liew2016scientific, SCHINTKE202482}. If we can accurately compute the mutual information between nodes, we can run the Chow-Liu algorithm to return an estimate of the electrical grid topology. Our solution therefore relies on calculating the mutual information between a pair of voltage distributions at distinct nodes. 

\begin{figure}
\centering\includegraphics[width=1.0\linewidth]{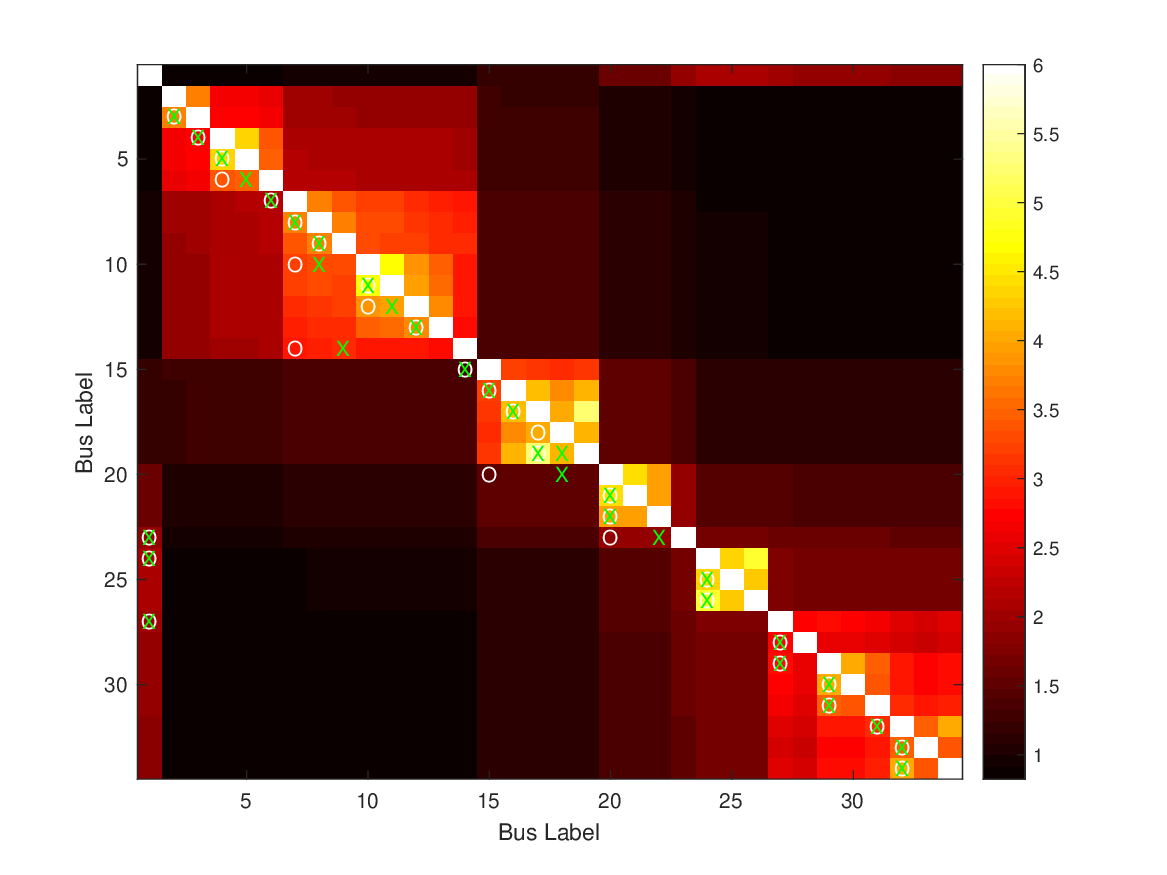}
\caption{The heatmap of mutual information is shown for data generated with GridLAB-D. The white circles represent true branches in the grid and the green crosses represent estimated branches based on the Chow-Liu approximation.}

\label{fg:heatmap1}
\end{figure}

We can further visualize how mutual information of pairs of nodes determines the branches we estimate. In Figure \ref{fg:heatmap1}, we can see the heatmap of mutual information for a 34 bus network. The self-information terms~\cite{stone2024information}, $MI(V_{i},V_{i})$, have the largest values in the heatmap but these values are not included in our calculations since in the Chow-Liu algorithm we look at the mutual information for pairs of distinct nodes. The heatmap shows that many of the largest MI values in the heatmap are marked with a green cross to show that the algorithm has estimated an edge between these two nodes. 

\subsection{Gaussian approximation}
Our program must therefore accurately compute the mutual information between voltage distributions at distinct pairs of nodes. The mutual information can be reformatted in terms of entropy terms as in Equation (\ref{eq:MI_entropy}) \cite{cover2012elements}. 

\begin{equation}
\label{eq:MI_entropy}
MI(X,Y) = H(X) + H(Y) - H(X,Y)
\end{equation}

The first method to compute mutual information is a Gaussian approximation method. To understand the method, we note that that the entropy of a discrete random variable $x$ is defined as in Equation (\ref{eq:entropy})~\cite{lemons2013student}.
\begin{equation}
\label{eq:entropy}
H(x) = - \sum_{i} p(x_{i})\log p(x_{i})
\end{equation}

\begin{figure}
\centering\includegraphics[width=0.6\linewidth]{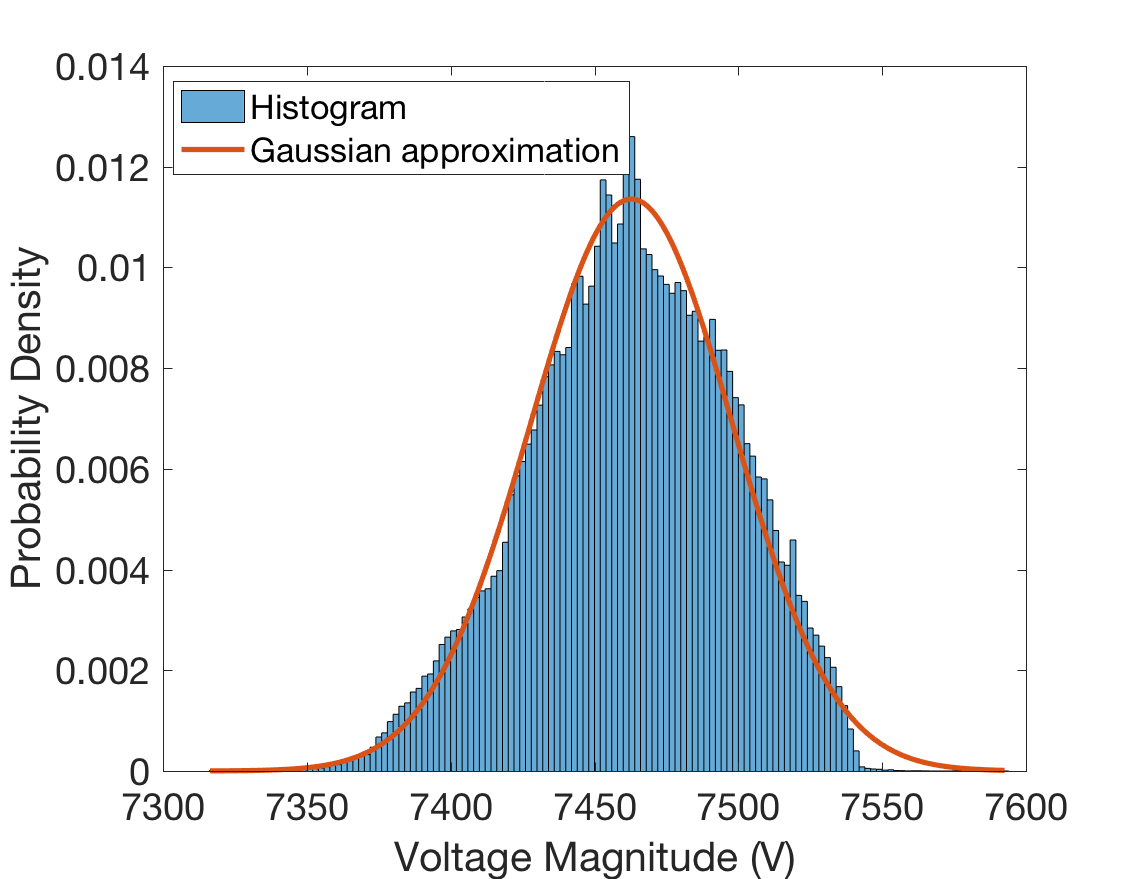}
\caption{Histogram and Gaussian model based on mean and standard deviation of voltage magnitude data at Node 18 from the SG1 dataset generated in GridLAB-D.} 
\label{fig:SG1_Histogram_Node18}
\end{figure}

Finding the entropy of any function can be approximated by discretizing the distribution and treating the observed frequency of an observation as the probability of the observation.
For large datasets, approximating the voltage distribution as a well known distribution can give a more computationally simple solution to finding the entropy and thereby mutual information than computing Equation (\ref{eq:entropy}). As often is the case for physical data, voltage magnitude data in the datasets generated using MATPOWER and GridLAB-D closely resembles a Gaussian distribution as seen in Figure \ref{fig:SG1_Histogram_Node18} for a 53-node network. The other datasets also show this trend. The entropy of a Gaussian function with mean $\mu$, covariance matrix $\Sigma$ and dimension $k$ is  displayed in Equation (\ref{eq:Gaussian_entropy}).

\begin{equation}
\label{eq:Gaussian_entropy}
H(X) = \frac{k}{2} (1+\ln {2\pi}) +  \frac{1}{2} ln(\det \Sigma)
\end{equation}

\subsection{Discrete mutual information}
As mentioned earlier, the more computationally expensive option for large datasets is to approximate probabilities for frequencies of occurrence in the datasets. The mutual information can also be calculated as in Equation (\ref{eq:MI_discrete}). The discrete estimate should yield the most accurate mutual information for large datasets since the frequency of observing a discrete value approaches the true probability for the discrete value~\cite{mackay2003information}. Note this process is quite slow for a network where nodes can take on many values and there is a large amount of data. Thus, we expect the discrete mutual information to give us the most accurate value but with the slowest performance. 

\begin{equation}
\label{eq:MI_discrete}
I(X,Y) = \sum_{x \in X}\sum_{y \in Y} p(x,y) \log \frac{p(x,y)}{p(x)p(y)}
\end{equation}

\subsection{JVHW Entropy Estimation}

Due to the computational expense of calculating the discrete MI of data, many research groups have explored computational short-cuts~\cite{walters2009estimation}. The Weissman group at Stanford University has developed minimax estimators to optimally estimate the entropy of a function~\cite{jiao2015minimax}. The estimators treat the smooth and non-smooth region of the function differently. A bias corrected MLE is applied to the smooth region of the function and the best polynomial fit to the non-smooth region. 
The JVHW estimator seeks to estimate a functional of entropy, $F_{\alpha}(P)$ defined as: 

\begin{equation}
\label{eq:entropy_functional}
F_{\alpha}(P) \triangleq \sum_{i =1}^{S} p_{i}^{\alpha}, \alpha > 0,
\end{equation}

where $\alpha > 0$ for a unknown discrete probability distribution $P = (p_{1}, p_{2}, ...,p_{S})$, with unknown support size $S$. Regarding estimation of the functional of entropy, the $L_{2}$ risk of an arbitrary estimator, $\hat{F}$, is defined as:

\begin{equation}
\label{eq:L_two_risk}
E_{P} (F(P) - \hat{F})^2.
\end{equation}

The $L_{2}$ risk is a function of both the unknown distribution $P$ and the estimator $\hat{F}$. The goal of the estimation is to minimize the $L_{2}$ risk. Since the distribution $P$ is unknown, the researchers could not directly minimize it, so instead, they try to minimize the maximum risk displayed in Equation (\ref{eq:JVHW}). The researchers showed similar accuracy to the discrete MI while observing a sizable improvement in computational performance.

\begin{equation}
\label{eq:JVHW}
\sup_{P \in M_{S}}  E_{P} (F(P) - \hat{F})^2
\end{equation}

\subsection{Datasets}

The data used in this article was generated using MATPOWER~\cite{zimmerman2010matpower}, a well-known grid simulation package for Matlab. The work described in this document seeks to reproduce the results of the Rajagopal group for an IEEE 123-bus network data generated with MATPOWER and extend results to data generated with GridLAB-D~\cite{chassin2008gridlab}, a software package developed by the Pacific Northwest National Laboratory to simulate power flow and control of electrical grids.  This manuscript presents the following listed contributions.

Using voltage data at different buses, our task is to estimate the connectivity of a sub-section or in some cases the entire topology of the electrical grid. For the scope of this document we focus on analyzing datasets where voltage magnitude data are available but voltage phase data are not. This document analyzes data that have been generated with two different software packages. The simulated data seek to represent data that could be gathered from smart-meters that take voltage measurements at constant time intervals. Data have been generated using MATPOWER to create IEEE standard networks, which are labeled as IEEE datasets. Using GridLAB-D, datasets have been generated to simulate realistic power profiles. The GridLAB-D datasets in some cases also have solar power introduced into the grid. The GridLAB-D datasets are labeled SG (non-solar) datasets when solar is not present in the grid and SG (solar) datasets when solar power is present in the grid.

In the IEEE datasets, we have one year of hourly voltage magnitude and phase data for each bus. Hence, there are $8760$ data points in the IEEE dataset for each bus. Similarly, the GridLAB-D datasets cover a year of data but have more precise resolution with a measurement taken every minute. This gives $525,610$ data points for each bus.  

The discrete MI method and JVHW method require discrete data as input. These estimators exhibit high computational complexity, scaling heavily with both the length of the time series and the number of discretization bins. For a network of $M$ nodes, computing all pairwise mutual information requires $O(M^2)$ operations. The methods can take on the order of days (e.g., approximately 29 hours for the 123-node system) when we discretize the data to higher resolution than 16-bit integers. We vary the discretization of datasets and look at the effect on algorithm performance in Section (\ref{S:3.6}). To reduce computation time, we run the bulk of our experiments with a 14-bit discretization for the discrete MI methods. The Gaussian MI, on the other hand, performs relatively fast (yielding network reconstructions in approximately 5 seconds for the same system) and performs well with 64-bit floating-point precision in Matlab, the highest floating-point resolution in Matlab.

There are four GridLAB-D datasets. Two datasets have 52 buses and are labeled SG1 (solar and non-solar). The other two GridLAB-D datasets have 272 buses and are labeled SG2 (solar and non-solar). The true graphs of SG1 and SG2 datasets are shown in the Appendix in Figures \ref{fg:SG1_true_graph} and \ref{fg:SG2_true_graph} respectively.

\subsection{Collapsing Redundant Data}
\begin{figure}
\centering\includegraphics[width=1.0\linewidth]{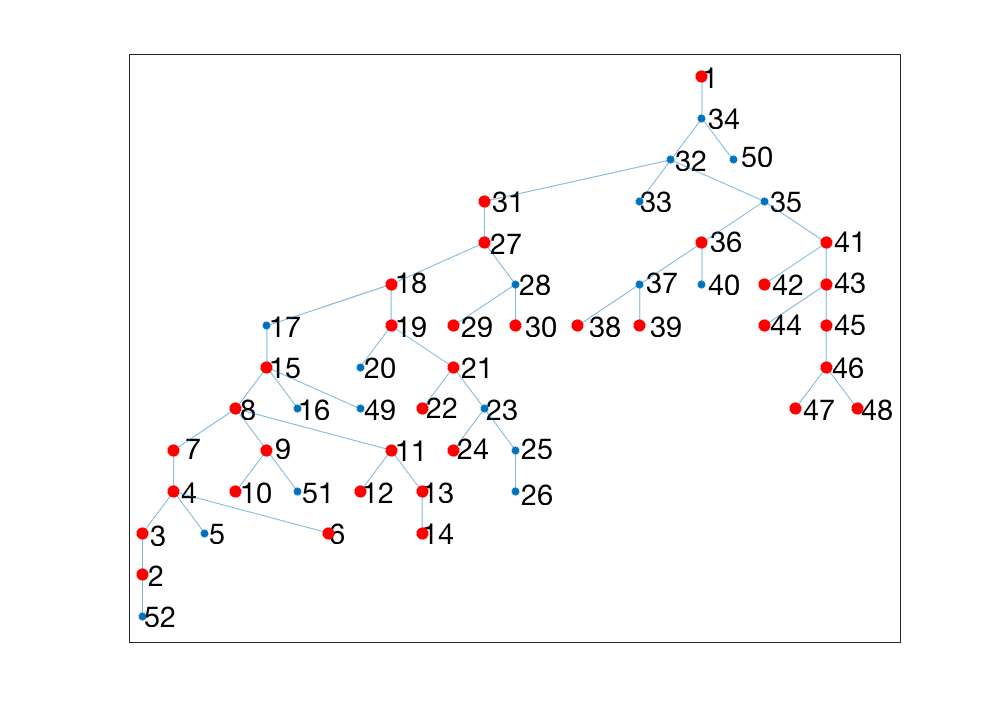}
\caption{The SG1 graph, based on a dataset generated in GridLAB-D, changes shape when redundant nodes are removed. The non-redundant nodes are marked red.}
\label{fg:SG1_collapsed}
\end{figure}

\begin{figure}
\centering\includegraphics[width=1.0\linewidth]{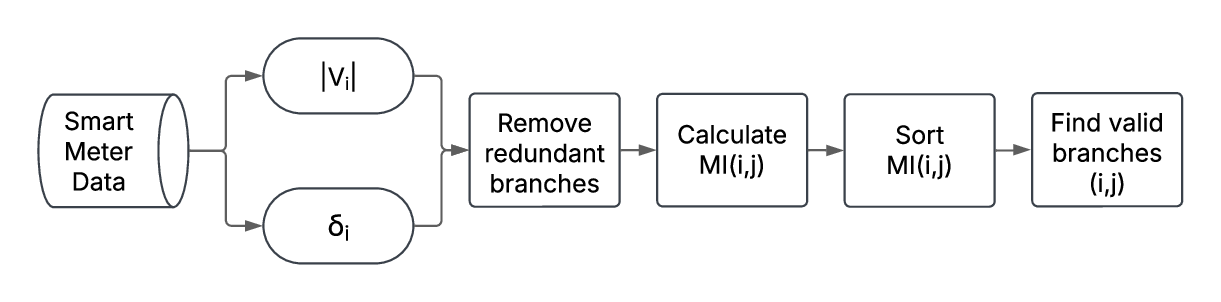}
\caption{The improved workflow removes redundant data before calculating mutual information. This added step greatly improves the performance of topology estimation.}
\label{fg:alg_flow_collapse}
\end{figure}

The datasets generated with GridLAB-D show nodes which contain the same data as other nodes. This is due to nodes being separated by fuses and switches or being in close proximity to one another. Collapsing nodes with the same data into a single node on the graph greatly improves estimation performance since the algorithm cannot distinguish nodes with identical data. We can see the redundant nodes in Figure~\ref{fg:SG1_collapsed}. With this step in mind, our program flow is adapted and displayed in Figure \ref{fg:alg_flow_collapse}.

\subsection{Performance Metrics}

The most important metric to measure performance is the percentage of correct branches found by the algorithm. We analyze datasets for which the true topology is known in this work. We estimate $m-1$ branches for a grid with $m$ buses/nodes. We can calculate the correct percentage of branches estimated by the algorithm using Equation (\ref{eq:SDR}). This metric is known as the successful detection rate is referred to hereafter as the SDR. 

\begin{equation}
\label{eq:SDR}
\text{SDR} \:= \: \frac{ \text{Number\ of\ branches\ correctly\ estimated}}{\text{Total\ number\ of\ branches}} * 100 \%
\end{equation}

A leaf node in the tree is a node which connects to only one other node. The leaf SDR is the proportion of leaf nodes that are correctly connected based solely on the mutual information. Therefore, we look at the mutual information for all pairs of buses with the leaf node in question. If the highest valued mutual information is at true connection, this leaf node has been correctly estimated based on MI. We employ this statistic to get a better sense of where the algorithm has difficulty correctly estimating the topology. 

\begin{equation}
\label{eq:leaf_SDR}
\text{Leaf SDR} \:= \: \frac{ \text{Number of leaf buses correctly estimated}}{\text{Total number of leaf buses}} * 100 \%
\end{equation}

\section{Results}

The results for the different MI methods for different datasets are shown below in Table \ref{tb:Optimal}. Note $\lvert\Delta  V \rvert $ refers to the change in voltage magnitude. Every method in the Table uses the change in voltage magnitude as the data for analysis. Note also, as discussed in the previous section, the Gaussian MI method uses 64-bit floating point precision while the JVHW and discrete MI methods use 14-bit integers as input for computational efficiency. The IEEE dataset is generated using MATPOWER with the condition that the power factor at each bus is chosen randomly. 

\begin{table}[h!]
\centering
\renewcommand{\arraystretch}{1.5}
\captionsetup{skip=10pt}
\begin{tabular}{p{4cm} p{2cm} p{4.8cm} p{1.5cm}}
\toprule
\textbf{Dataset} & \textbf{Number of Nodes} & \textbf{MI Method} & \textbf{SDR (\%)} \\
\midrule
SG1 & 34 & Gaussian/JVHW/Discrete $ \lvert \Delta V \rvert $  & 100 \\
SG1 (solar) & 34 & Gaussian/JVHW/Discrete $ \lvert \Delta V \rvert $& 100 \\
SG2 & 220 & Gaussian, $ \lvert \Delta V \rvert $  & 94.98 \\
SG2 (solar) & 220 & Gaussian/Discrete, $ \lvert \Delta V \rvert $  & 94.98 \\
IEEE Random Power Factor & 123 & Gaussian/JVHW/Discrete, $ \lvert \Delta V \rvert $ & 100 \\
\bottomrule
\end{tabular}
\caption{The optimal method and performance for year-long voltage magnitude datasets. Note number of nodes here refers to the number of non-redundant nodes in the network.}
\label{tb:Optimal}
\end{table}

\subsection{Algorithm Performance with Change in Voltage Magnitude}

We investigate the effect on the SDR when using the voltage magnitude of a bus compared to the change in voltage magnitude of a bus as the fundamental variable for the Chow-Liu algorithm. The results of using both the voltage magnitude and the change in voltage magnitude are shown in Table \ref{tb:Deriv}. Using $\lvert \Delta V \rvert$ as the fundamental variable shows superior performance for both the GridLAB-D and IEEE datasets for all methods of computing mutual information. With this result in mind, we use $\lvert \Delta V \rvert$ almost exclusively as our variable of analysis for GridLAB-D datasets in the sections that follow. 

\begin{table}[h!]
\centering
\renewcommand{\arraystretch}{1.4}
\captionsetup{skip=10pt}
\begin{tabular}{l l l}
\toprule
\textbf{Dataset} & \textbf{MI Method} & \textbf{SDR (\%)} \\
\midrule
SG1 non-solar & Gaussian $ \lvert \Delta V \rvert $  & 100 \\
SG1 non-solar & Gaussian   & 60.61 \\
SG1 non-solar & Discrete $ \lvert \Delta V \rvert $  & 100 \\
SG1 non-solar & Discrete   & 63.64 \\
SG1 solar & Gaussian $ \lvert \Delta V \rvert $  & 100 \\
SG1 solar & Gaussian   & 60.61 \\
SG1 solar & Discrete $ \lvert \Delta V \rvert $  & 100 \\
SG1 solar & Discrete   & 66.67 \\
IEEE 123-Node Random Power Factor & Gaussian, $\lvert \Delta V \rvert $ & 100\\
IEEE 123-Node Random Power Factor & Gaussian & 95.85 \\
\bottomrule
\end{tabular}
\caption{The effect on the SDR is shown for when $ \lvert \Delta V \rvert $ is used as the fundamental variable.}
\label{tb:Deriv}
\end{table}

\subsection{Gaussian Approximation of Data}

To get a better sense of the nature of the different datasets, we may examine the Kullback-Leibler (KL) divergence~\cite{kullback1951information} between the voltage magnitude at one node and a Gaussian fit based on the data's mean and variance. The KL divergence between two probability distributions, $p$ and $q$, is a measure of similarity of the distributions. The formula for the KL divergence is shown in Equation~\ref{eq:KL_div}. Note, unlike mutual information, the KL divergence is not symmetric. 

\begin{equation}
\label{eq:KL_div}
D_{KL}(p || q) = \sum_{x \in X}  p(x) \log{\frac{p(x)}{q(x)}}
\end{equation}

To compute the KL-divergence the data at each node are binned into a thousand different bins. The Gaussian approximation for the data is also computed at the same thousand voltage magnitude (or change in voltage magnitude) values as the binned data. The KL-divergence is then computed by using both sets of data and Equation~\ref{eq:KL_div}. The histogram and Gaussian approximation for the change in voltage magnitude at Node 33 of the SG1 dataset is shown in the Appendix in Figure \ref{fg:sg1_node_33_hist}.

The mean, min, max, and standard deviation of the KL divergence between the change in voltage magnitude histogram and the Gaussian approximation for each dataset is shown in Table (\ref{tb:mag_KL_sg}). Note, the KL divergence is computed at each node in the dataset and the mean refers to the mean KL divergence across all nodes in a specific dataset. The SG1 non-solar and SG2 solar/non-solar datasets all have similar mean KL divergence at around 7 bits. The SG1-solar shows a slightly worse fit at above 8 bits.

\begin{table}[h!]
\centering
\renewcommand{\arraystretch}{1.5}
\captionsetup{skip=10pt}
\begin{tabular}{l c c c c}
\toprule
\textbf{Dataset} & \textbf{Max (bits)} & \textbf{Min (bits)} & \textbf{Mean (bits)} & \textbf{Std (bits)} \\
\midrule
SG1 non-solar & 12.44&	5.46&	7.08&	2.05\\
SG1 solar & 13.00&	6.80&	8.17&	1.84 \\
SG2 non-solar & 14.53	&6.08	&7.26	&1.54 \\
SG2 solar & 14.54&	6.07&	7.13&	1.56 \\
\bottomrule
\end{tabular}
\caption{Statistics summary of the KL divergence between voltage magnitude data and the Gaussian approximation for different datasets.}
\label{tb:mag_KL_sg}
\end{table}

\subsection{Reducing Data Requirements for Estimation}

Most smart meters deployed in Canada sample at the most frequent rate of 15 minutes~\cite{lee2021data}. In California, older models sample once every half hour or once every hour in residential buildings~\cite{li2021characterizing}. In Norway, the UK and France, this sampling happens every thirty minutes~\cite{lee2021data}. We down-sample the datasets generated in GridLAB-D from one sample every 1 minute to commercially available resolutions such as 5, 15, 30, or 60 minutes per data point and observe the effect on the SDR.

\begin{figure}[h!]
\centering\includegraphics[width=1.0\linewidth]{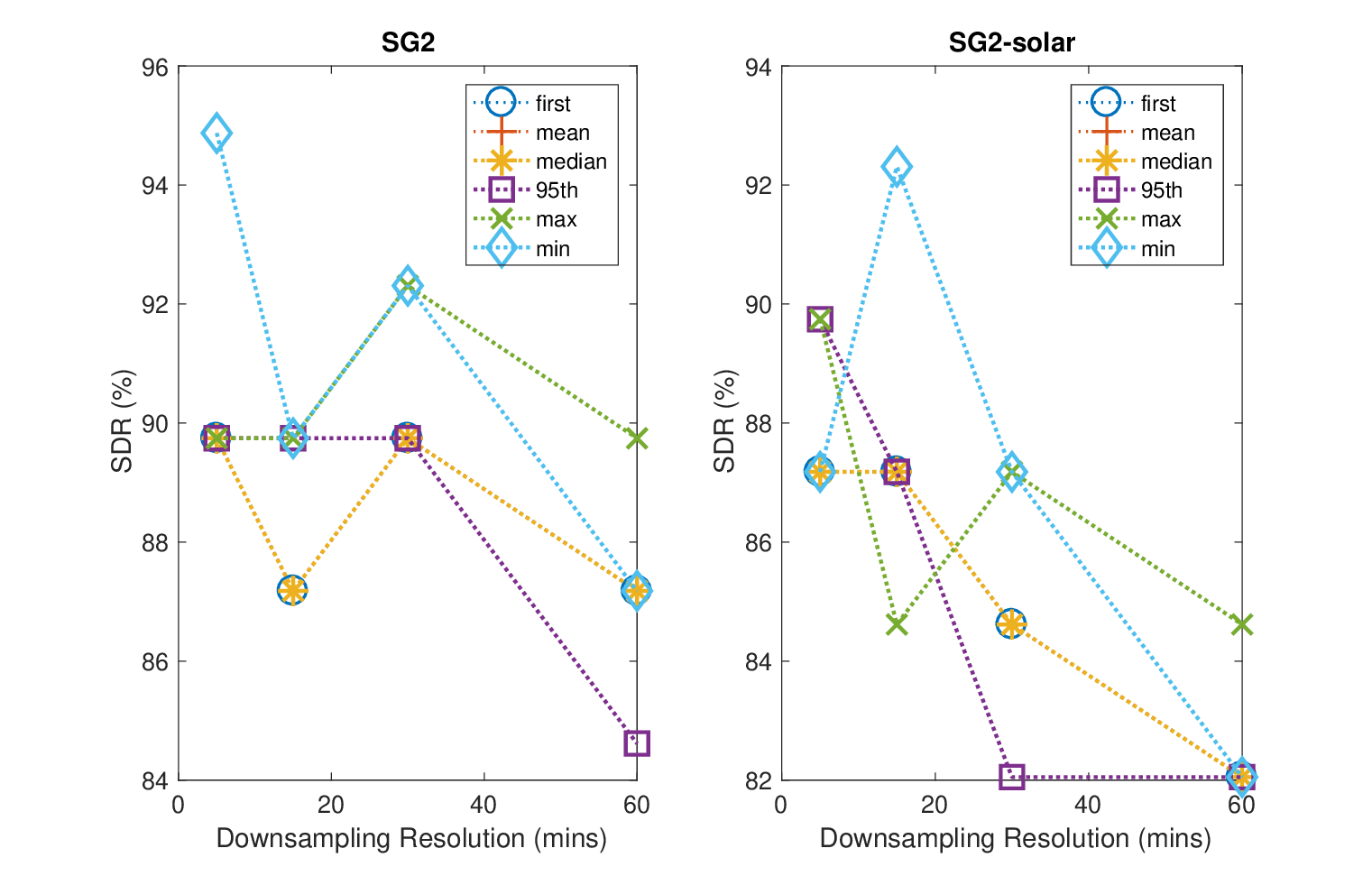}
\caption{The SG2 dataset is down-sampled from the original dataset containing one sample per minute to 1 sample per 5, 15, 30 and 60 minutes and then the change in voltage magnitude of the dataset is taken before estimation. Different down-sampling methods are compared on the full year dataset.}
\label{fg:SG2_dsample}
\end{figure}

For a fixed downsampling frequency we compare sampling the last or the first data-point of each period and compare performance in terms of SDR and leaf SDR to taking the mean, the median, the $95\%$ quantile and the max of each period of the data. The graph for the SG1 dataset in the Appendix Figure \ref{fg:SG1_dsample} does not show a clear favorite across down-sampling resolutions for different down-sampling methods. The graph for the SG2 dataset in Figure \ref{fg:SG2_dsample} also shows no clear winner. It appears the downsampling method where the first value from each interval is sampled, labeled \textit{first} in \ref{fg:SG1_dsample}, should be sufficient in the following experiments. The \textit{first} downsampling method also has the additional benefit of requiring simple circuitry to sample one value at a constant interval for smart meters.

\begin{figure}[h!]
\centering\includegraphics[width=1.0\linewidth]{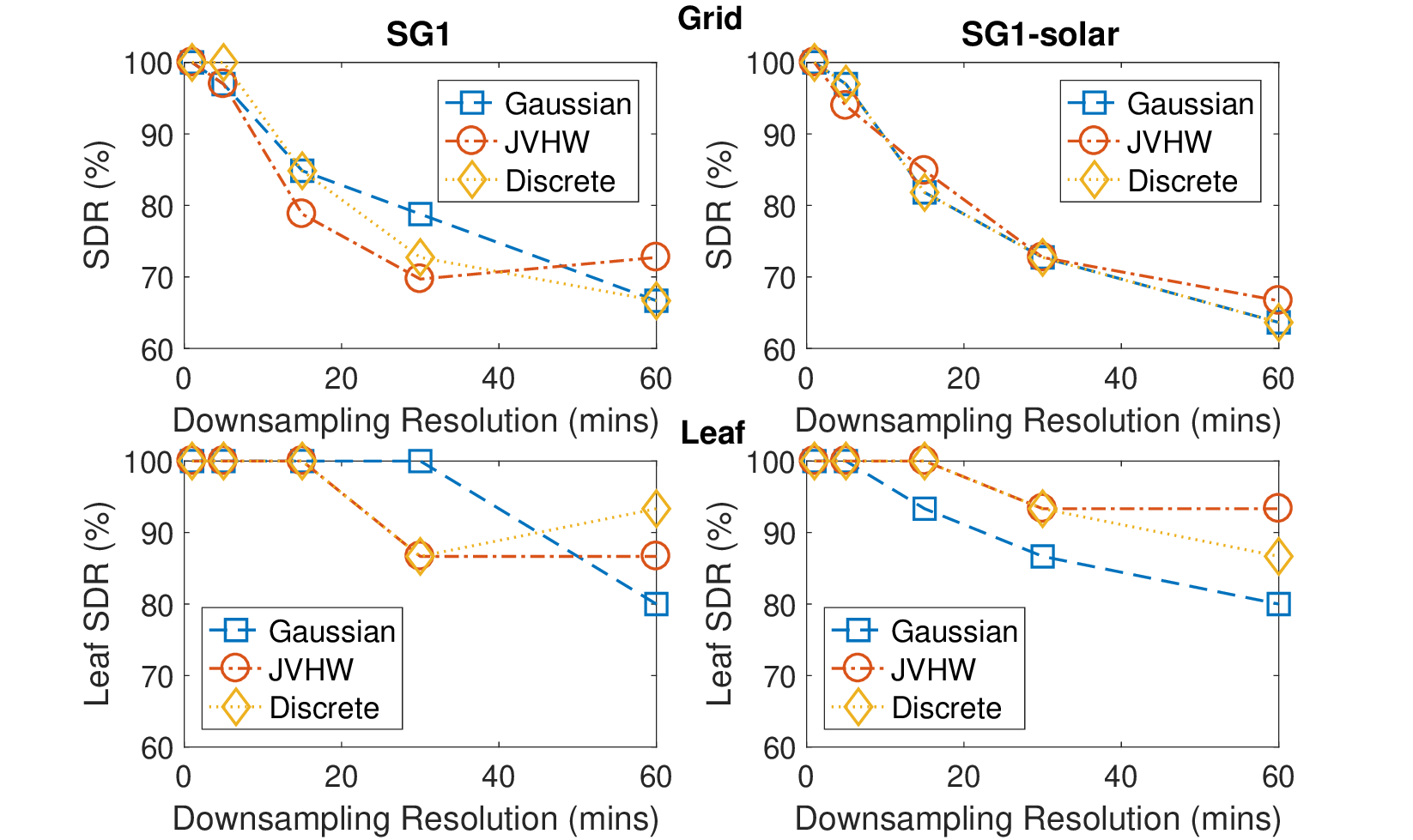}
\caption{The SG1 dataset is down-sampled from the original dataset containing one sample per minute to 1 sample per 5, 15, 30 and 60 minutes and then the change in voltage magnitude of the dataset is taken before estimation.}
\label{fg:SG1_res}
\end{figure}

We can also investigate the effect of downsampling when changing the method to calculate the mutual information. For the SG1 (non-solar) dataset, in Figure~\ref{fg:SG1_res} all the MI methods perform quite similarly except at a few data-points. The JVHW method performs significantly worse for the 15 minute (approximately 6\% worse) resolution compared to other methods while the Gaussian method performs significantly better than the two other discrete and JVHW methods (greater than 6\%) at 30 min resolutions. At a 60 minute resolution, the JVHW method performs approximately 5\% percent better than the other methods. For the SG1 (solar) dataset, all the MI methods perform quite similarly. SDR performance drop off as a function of resolution aligns closely between the two SG1 datasets. Moreover, the performance drop at 15 minutes and lower resolutions aligns closely with the performance drop we saw for variable change in voltage magnitude step sizes of 15 minutes or greater in the previous section. 

With regards to the SG1 leaf SDR in Figure \ref{fg:SG1_res}, all methods show 100\% leaf node SDR for 1 min, 5 min and 15 min resolutions. In contrast to what we see with the SG1 (solar) dataset, the SG1 (non-solar) shows that the Gaussian method has the best performance for 30 min resolution with a perfect 100\% leaf SDR.  For SG1 (solar), the JVHW and discrete methods show 100\% leaf node SDR for 1 min, 5 min and 15 min resolutions. For SG1 (solar), leaf SDR performance drops off by approximately 6\% from 15 min to 30 min resolution. 

Further experiments are needed to explain the strange leaf node SDR behavior of the discrete MI method in SG1 where the performance increases significantly from 30 minute resolution to 60 minute resolution. We can take a look at Histograms and Gaussian fits in the Appendix Figure \ref{fg:SG1_hist_res} to understand the effect of resolution on the data. The histograms show that the change in voltage magnitude becomes more spread out as the resolution increases. We are not surprised by this behavior of larger changes between time intervals for larger time intervals. There does not appear to be any information in the histograms to explain the increase in leaf SDR by the discrete MI method from 30 minute to 60 minute resolutions.

To get a better sense of how downsampling affects the algorithm we can visualize Node 7 in the SG1 non-solar dataset and its associated mutual information rankings as a heatmap in the appendix Figure \ref{fg:SG1_MI_dsample_heatmap}. The heatmap ranks the mutual information pairings with Node 7 as a function of downsampling resolution. Note, the pairing ranked number 1 represents the highest mutual information pairing. The self-information, Node 7 paired with Node 7 is, as we expect, ranked the highest value pairing throughout downsampling. We see that Nodes 2 and 3 are ranked the same throughout downsampling. The lower ranked nodes are more important in terms of algorithm performance since the maximum spanning tree is selected choosing higher value mutual information pairs corresponding to low rank pairings in this analysis. Nodes 4,5,6 change ranking as the downsampling resolution increases to sixty minutes per sample. Otherwise the other rankings stay largely the same.

The same analysis is performed on Node 23 in Figure \ref{fg:SG1_MI_dsample_node23} of SG1 non-solar for the Gaussian MI method using the change in voltage magnitude data. The Figure shows that rankings 2-5 do not change through the downsampling process. We see ranking numbers 6,8 and 14 are different as we progress from the original SG1 dataset to the 60 minute/sample downsampled version. Considering there is such a large effect on the SDR in downsampling the SG1 dataset, the fact that there are not many mutual information ranking changes is surprising.

The mutual information ranking analysis is also performed on the SG1 solar dataset using the Gaussian mutual information on the change in voltage magnitude data. Node 23 is shown in the Appendix Figure \ref{fg:SG1_solar_MI_dsample_node23}. The Figure shows very similar ranking changes to SG1 non-solar Node 23 that was previously examined. Ranking numbers 2-5 are consistent throughout downsampling while numbers 6,8,14 and 15 change from the original dataset to the sixty minutes/sample dataset. The great drop in SDR from the original SG1 solar/non-solar datasets to the downsampled versions suggests in this analysis that very small perturbations in mutual information rankings can greatly affect the estimated tree and resulting SDR.

In the Appendix Figure \ref{fg:SG2_res}, for the SG2 (solar) and (non-solar) data, the Gaussian method performs the best for 1 and 5 min resolutions in terms of SDR. The Gaussian method performs significantly better than other MI methods at 60 min resolution (approximately 3\% greater than the JVHW method and approximately 6\% greater than discrete method) in terms of SDR. 

Moreover, the SG2 (solar) data shows that the JVHW method improves slightly at 5 min resolution compared to 1 min, which goes against our expectations for lower SDR with lower resolution. It's also odd that the SDR for the JVHW method improves at 30 min resolution compared to 15 min resolution. The Gaussian method is the best at 1 min resolution. The Gaussian method has less than a 0.5\% percent greater SDR at 1 min res compared to discrete and approximately 2\% percent improvement at 5 min resolution but no improvement compared to the discrete method at 15 min resolution. The Gaussian method at 60 min resolution shows the best leaf SDR in Figure \ref{fg:SG2_res}. We see an approximately 2\% increase from 1 minute to 60 minute resolution for the Gaussian method. The leaf node SDR is more than 7\% lower than the regular SDR for all MI methods for the SG2 (solar) dataset. This suggests the great majority of estimation mistakes for the grid topology are made in estimating leaf branches.

\subsubsection{Data Length Analysis} 

We also investigate what time window of data is required to estimate the grid topology with accuracy. We refer to a time window of data as a length of data (e.g., 100 consecutive days of data). We can look at the effect on SDR for the SG1 (non-solar) dataset for length sizes smaller than one year in Figure \ref{fg:SG1_lens}. The JVHW and discrete methods perform significantly better than the Gaussian method for smaller lengths of data. Both the JVHW and discrete MI methods show greater than 98\% SDR at 60 day lengths and the JVHW method shows greater than 97\% SDR at 30 day lengths.

\begin{figure}[h!]
\centering\includegraphics[width=0.6\linewidth]{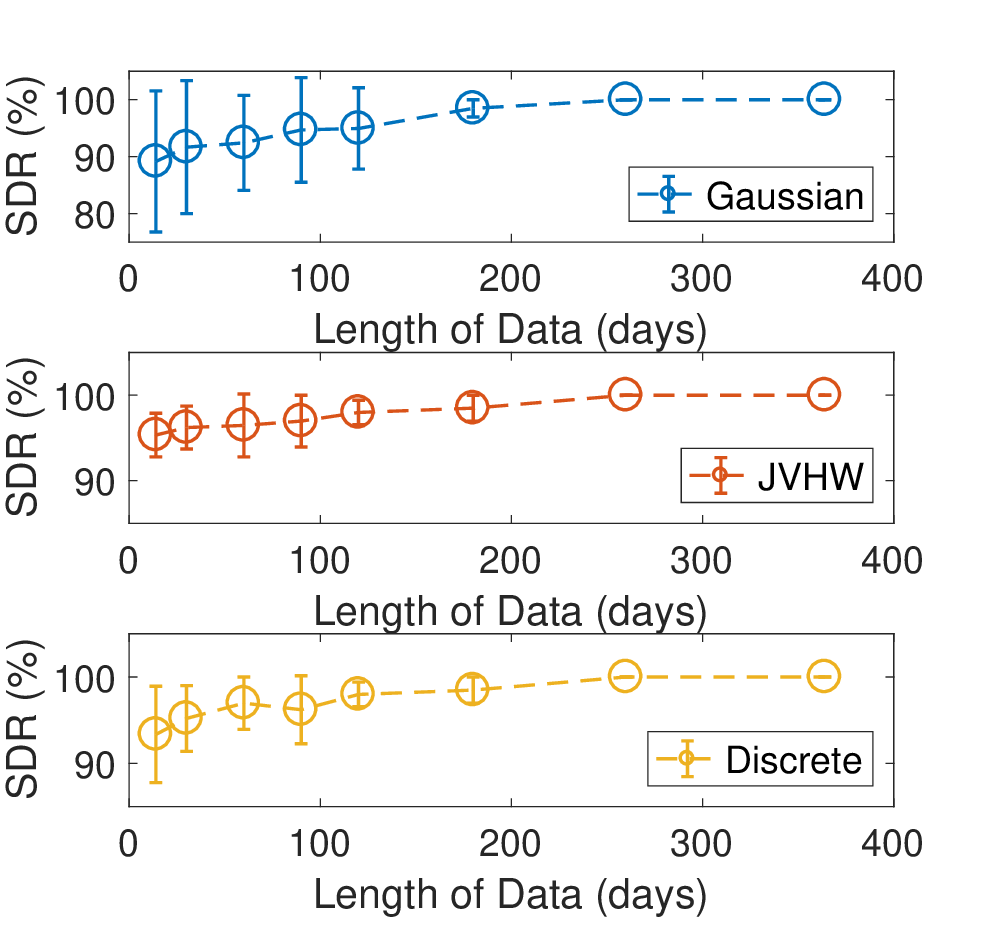}
\caption{Lengths of consecutive days of data are taken of the SG1 (solar) dataset before estimation.}
\label{fg:SG1_lens_solar}
\end{figure}

Figure \ref{fg:SG1_lens_solar} shows the effect of length size of data on SDR for the SG1 (solar) dataset. All the MI methods perform similarly to one another. The Gaussian, JVHW and discrete methods show greater than 98\% SDR for 180 days and up. In this regard, the SG1 (solar) dataset requires a greater length of data for accurate estimation than the SG1 dataset.

The effect on the leaf SDR from length size is shown in the Appendix Figure \ref{fg:SG1_lens_leaf} for the SG1 (non-solar) dataset. The JVHW and discrete MI methods outperform the Gaussian method. The leaf node mean SDR is greater than 99.5\% for the discrete MI methods for a length of data of seven days. A similar plot is shown for the SG1 (solar) dataset in the Appendix Figure \ref{fg:SG1solar_lens_leaf}. Less data, a 4 day length, are required for greater than 99.5\% SDR for the discrete MI methods for the SG1 (solar) dataset compared to the non-solar data.

The effect on the SDR of different length sizes of data is shown for the SG2 (non-solar) dataset in the Appendix Figure \ref{fg:SG2_lens}. With only two weeks of data all MI methods are less than 0.8\% away from the full year's SDR. The Gaussian method performs the best of all MI methods for all length sizes. The Gaussian method is within 1\% of the full length of data with only one day’s worth of data (at 1 min resolution).

Figure \ref{fg:SG2solar_lens} in the Appendix shows the effect on SDR of different data length sizes for the SG2 (solar) dataset. With only two weeks of data the Gaussian and JVHW methods are less than 0.6\% away from the full years length of data SDR. At two weeks the discrete method is less than 1.5\% away from the full length size SDR. The Gaussian method performs the best of all MI methods for all length sizes. The Gaussian method is within 1.5\% of the full length of data with only one day’s worth of data (at 1 min resolution).

The effect on the leaf SDR of different length sizes is shown in the Appendix Figure \ref{fg:SG2_lens_leaf} and Figure \ref{fg:SG2solar_lens_leaf} for the SG2 (non-solar) and SG2 (solar) datasets respectively. There is more variance in the leaf node SDR than the regular SDR for the SG2 datasets. The leaf node SDR is lower than the regular SDR (we saw this trend already in the resolution analysis) for the SG2 datasets. The Gaussian performs the best across all length sizes for the SG2 datasets.

\begin{figure}[h!]
\centering\includegraphics[width=0.6\linewidth]{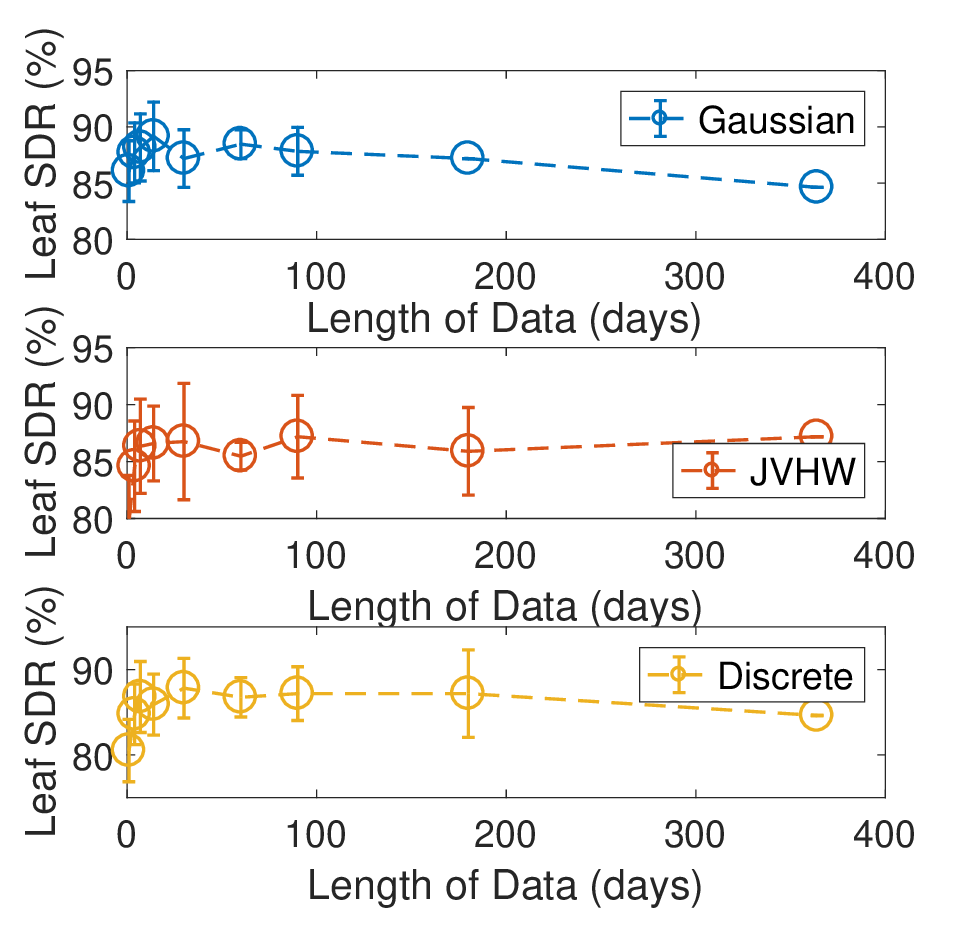}
\caption{Different lengths of consecutive days of data are taken of the SG2 (solar) dataset before estimation.}
\label{fg:SG2solar_lens_leaf}
\end{figure}

\begin{figure}[h!]
\centering\includegraphics[width=0.7\linewidth]{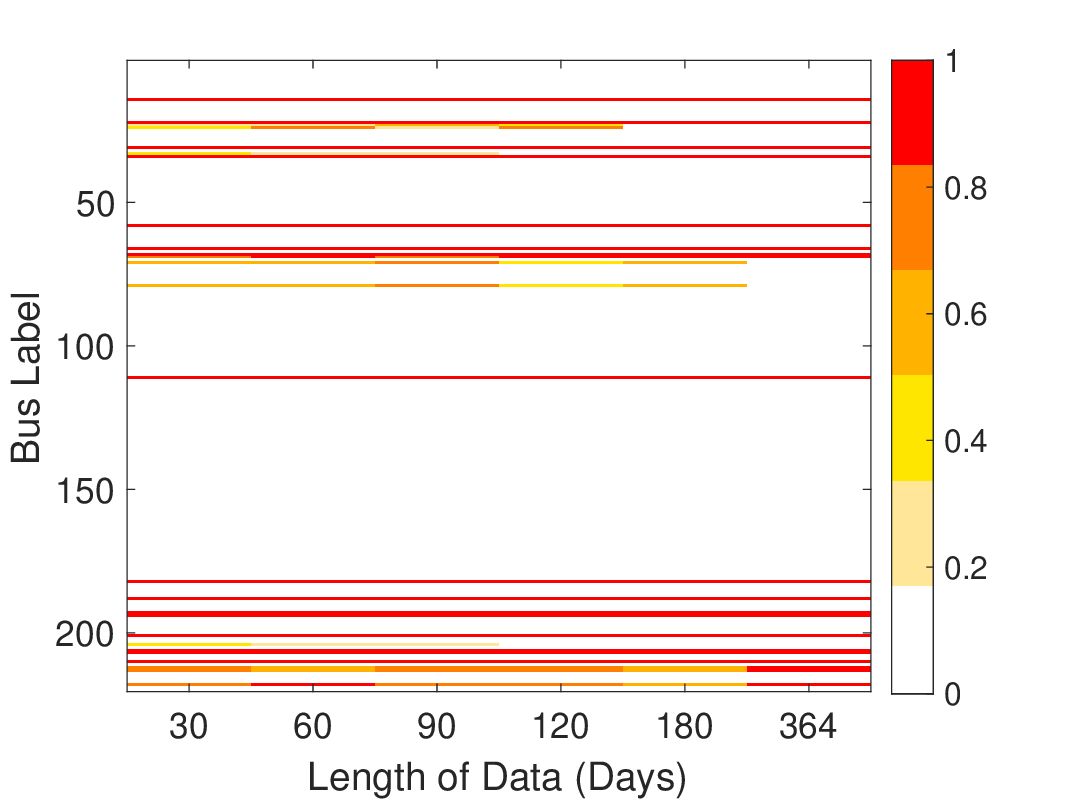}
\caption{Heatmap representing the frequency that a node is estimated to be connected to an incorrect branch. Analysis for SG2 (solar) using the Gaussian MI method on the change in voltage magnitude. The error frequency is a function of the lengths of days of data used in the analysis. Lengths of consecutive days of data are taken of the SG2 (solar) dataset before estimation. Note the 364 days of data is one-shot since only one non-overlapping length of data can be found in the year long dataset. }
\label{fg:SG2_solar_err_freq_heatmap_gaussian}
\end{figure}

The SG2 solar/non-solar datasets show a rather stagnant SDR from a 30 day length of data to a full year. We expect the SDR to increase with a larger dataset. Of interest is whether the same nodes are continuously wrongly predicted as the length of data is increased. The error frequency is defined to be the frequency that a node is estimated to be connected with an incorrect branch. For the 30 day length of data the error frequency is defined as the average over all of the consecutive lengths. For the 364 day length the error frequency is not averaged since there is only one non-overlapping length in the year long dataset. Note, if a node is estimated to have three branches and one is connected incorrectly  for all thirty consecutive day lengths, then the error frequency will be equal to one. The error frequency heatmap for the SG2 (solar) dataset using the Gaussian mutual information on the change in voltage magnitude is shown in Figure \ref{fg:SG2_solar_err_freq_heatmap_gaussian}. The heatmap shows that the same nodes are consistently connected with a branch. This suggests that the algorithm is consistently troubled by the same nodes regardless of data length for the SG2 (solar) incremental change in voltage magnitude data. 

The heatmap error frequency analysis has shown that largely the same set of nodes are predicted incorrectly by the algorithm for the SG2 solar/non-solar datasets. This suggests that reason why the algorithm does not show an increase in SDR over larger data lengths for SG2 solar/non-solar datasets is a result of a core set of nodes. Further study is required to determine if this core set of nodes hold unique properties within the dataset.

\subsection{Data Precision}
\label{S:3.6}

With massive amounts of data sent from commercial smart meters daily, data-compression can be a cost-reducing technique for utility companies. We can reduce data requirements of the voltage magnitude data before estimating the grid topology by reducing the data precision that is sent to the utility companies.

\subsubsection{Number of Bits of Data}

To observe the effect of quantization on the algorithm performance, we discretize datasets generated in GridLAB-D to as low as 4 bits and as high as 16 bits. Sixteen bits is chosen as the upper limit since the JVHW and discrete become very slow (greater than 40 minutes run-time on the Stanford barley servers) at higher resolution discretizations. We discretize by dividing the maximum and minimum voltage magnitude into equally sized $2^{N-1}$ intervals. The incremental change of voltage is used for quantization since we observed that discretizing before the incremental change in voltage is taken showed poorer performance. The histograms of two nodes are shown from the SG1 dataset (non-solar) in the Appendix Figure \ref{fg:SG1_hist_num_bits} for different levels of discretization.

\begin{figure}[h!]
\centering\includegraphics[width=0.7\linewidth]{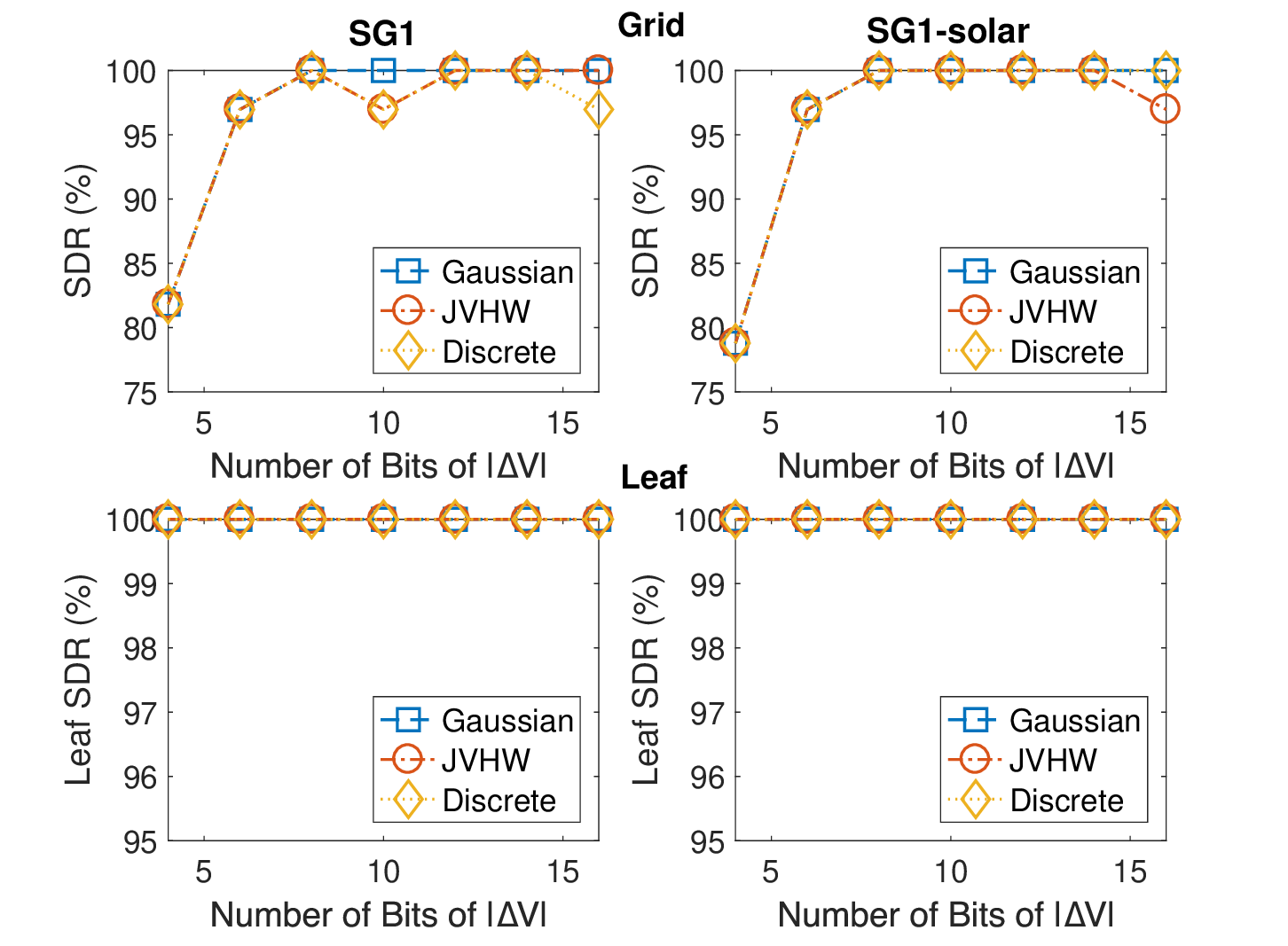}
\caption{The SG1 dataset is discretized to a certain amount of bits.}
\label{fg:SG1_num_bits}
\end{figure}

For the full lengths of data we want to see how the number of bits affects the SDR and leaf SDR. Figure \ref{fg:SG1_num_bits} shows that only 4 bits are required for perfect leaf SDR estimation for SG1 datasets. All the MI methods perform perfectly at 6, 12, 14 bits for SG1 and perfectly for 6, 8, 12, and 14 bits for SG1 (solar). The SG1 (non-solar) dataset shows the discrete method performing worse at 16 bits compared to 14 bits. The SG1 (solar) data-set shows the JVHW method performing worse for 16 bits compared to 14 bits. We expect increased performance with higher digitization resolution. 

\begin{figure}[h!]
\centering\includegraphics[width=0.7\linewidth]{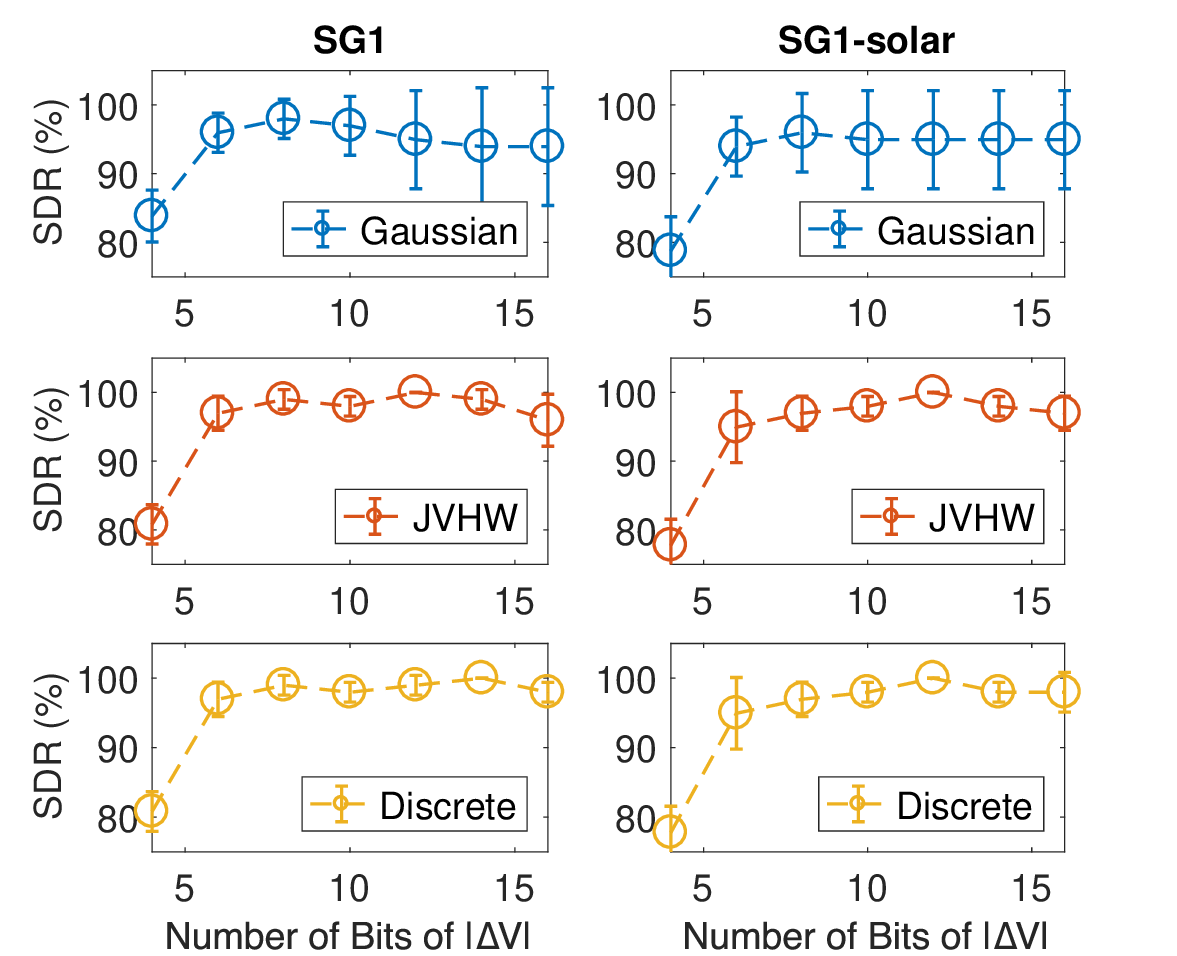}
\caption{The SG1 dataset is discretized to a certain amount of bits. Here a length size of 120 days is used. The lengths of data are taken consecutively in the year-long dataset, meaning three sets of 120 days are used in the analysis.}
\label{fg:SG1_num_bits_lens}
\end{figure}

To get a clearer picture of whether 16 bits performs worse than 14 bits for the discrete MI methods, we perform the same analysis with three sets of consecutive non-overlapping data, each a 120 day length of data of the full year dataset. The results for the SG1 datasets are shown in Figure \ref{fg:SG1_num_bits_lens}. The results do not clearly show the expected trend of increased SDR with greater digitization resolution nor reduced variance with increasing resolution. The leaf SDR for 120 day lengths sizes of data for the SG1 datasets is shown in Figure \ref{fg:SG1_num_bits_lens_leaf}. This graph shows perfect leaf SDR performance for 6-bits and over.

The SG2 dataset offers another glimpse into the effect of data precision on the algorithm's performance. Figure \ref{fg:SG2_num_bits} shows us there is not much impact in the SDR at resolutions greater than 8-bit. The leaf SDR, however, shows a greater than 5\% increase from 10-bit to 16-bit for all methods. The leaf SDR results suggest that 16-bit should be used for the discrete MI methods since they show a greater than 5\% increase in leaf SDR from 14-bit to 16-bit. 

\begin{figure}[h!]
\centering\includegraphics[width=0.7\linewidth]{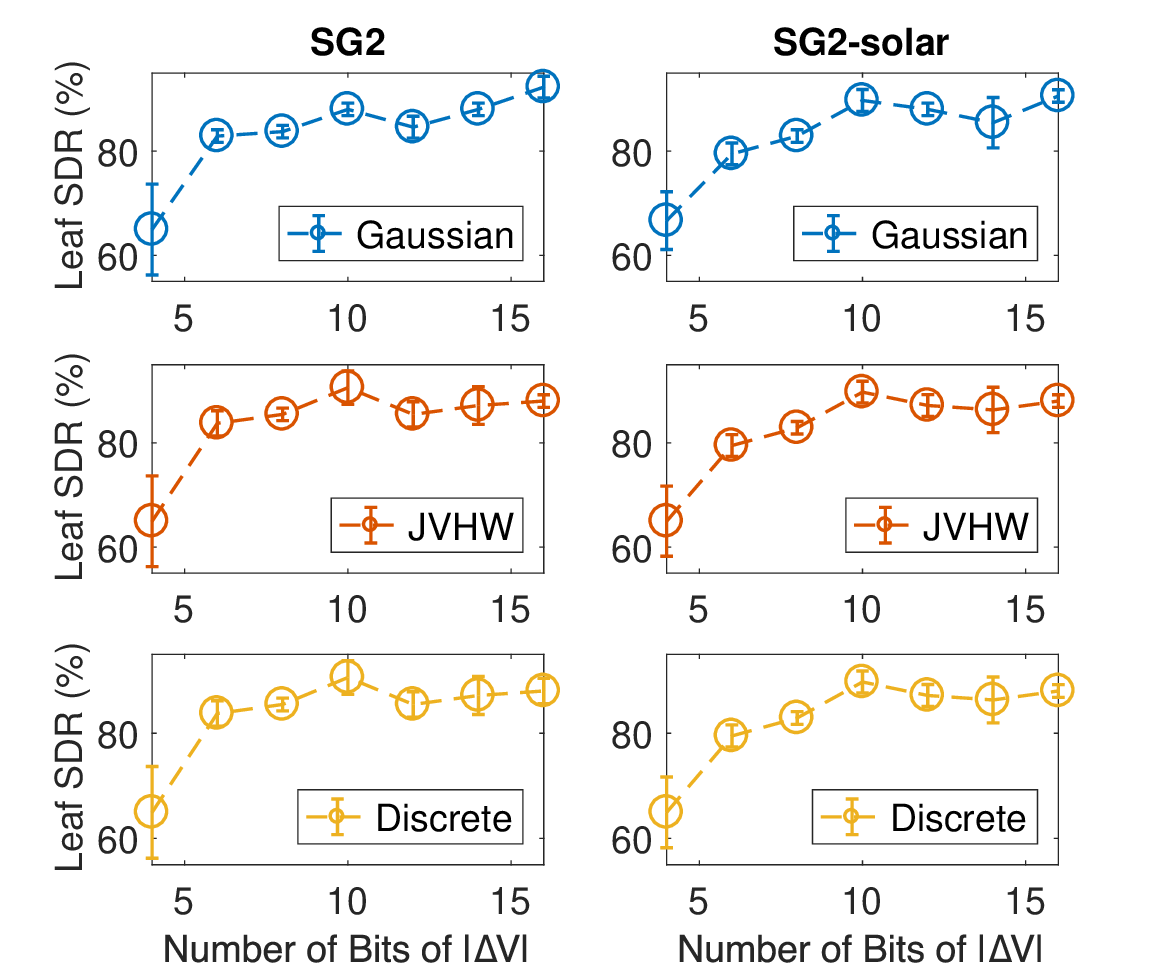}
\caption{The SG2 dataset is discretized to a certain amount of bits. Here a length of 120 days is used. The lengths of data are taken consecutively in the year-long dataset, meaning three sets of 120 days are used in the analysis.}
\label{fg:SG2_num_bits_lens_leaf}
\end{figure}

The results so far discussed are one-shot, meaning they represent only one data point. Let's take a look at a 120 day length size of data for SG2. Appendix Figure \ref{fg:SG2_num_bits_lens} shows that each MI method does not show much improvement in SDR at resolutions greater than 10-bits. Figure \ref{fg:SG2_num_bits_lens_leaf} shows that all the MI methods have near identical performance for the leaf SDR. The Gaussian method shows an almost 1\% greater performance than the other methods at 16-bits. Each method has a few percentage points increase in performance from 14-bit to 16-bit. This graph would suggest 16-bit discrete and JVHW methods will grant the best performance. 

\subsubsection{Rounding Data to a Significant Digit}
\begin{figure}[h!]
\centering\includegraphics[width=0.7\linewidth]{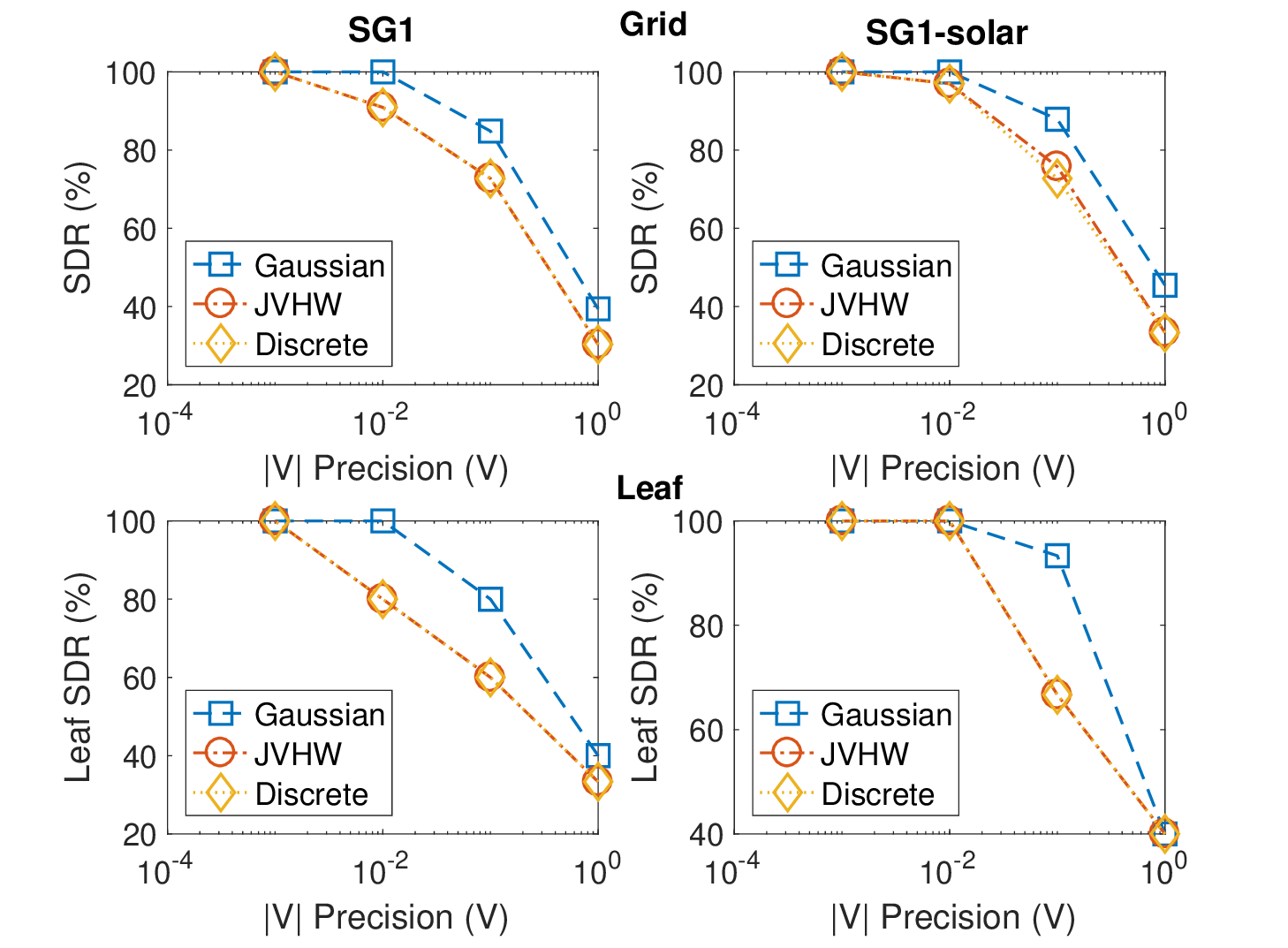}
\caption{The SG1 datasets are rounded to the nearest significant digit before analysis.}
\label{fg:SG1_sig_digs}
\end{figure}

Besides discretizing our data to a certain amount of bits we can also round our data to a significant digit of voltage magnitude data. Understanding these significant digit thresholds establishes a theoretical limit on how much data compression can be applied before structural inference degrades. The data are rounded to the most significant digit of analysis prior to discretizing the data. This method has similarities to the number of bits analysis, however, the number of bits analysis rounds the continuous data to the closest integer bit value after scaling, whereas the following analysis simply rounds to the closest integer without scaling.  
In Figure \ref{fg:SG1_sig_digs}, we can see the effect of using SG1 data to the nearest significant digit on the SDR. Note that the Gaussian method is rounded before analysis to a significant digit. The other MI methods have data multiplied by a base ten number and then rounded to nearest integer before discrete MI method analysis (this is done since the JVHW and discrete MI methods only accept positive integers as input due to their implementation in this work). The Gaussian method surprisingly outperforms the other MI methods. For SG1, it’s clear that only millivolts (mV) of data are needed for all MI methods. The Gaussian method outperforms the other method needing 10’s of mV of accuracy. For SG1 (solar), the leaf SDR shows all methods need only 10’s of millivolts of data to perform optimally.

\begin{figure}[h!]
\centering\includegraphics[width=0.7\linewidth]{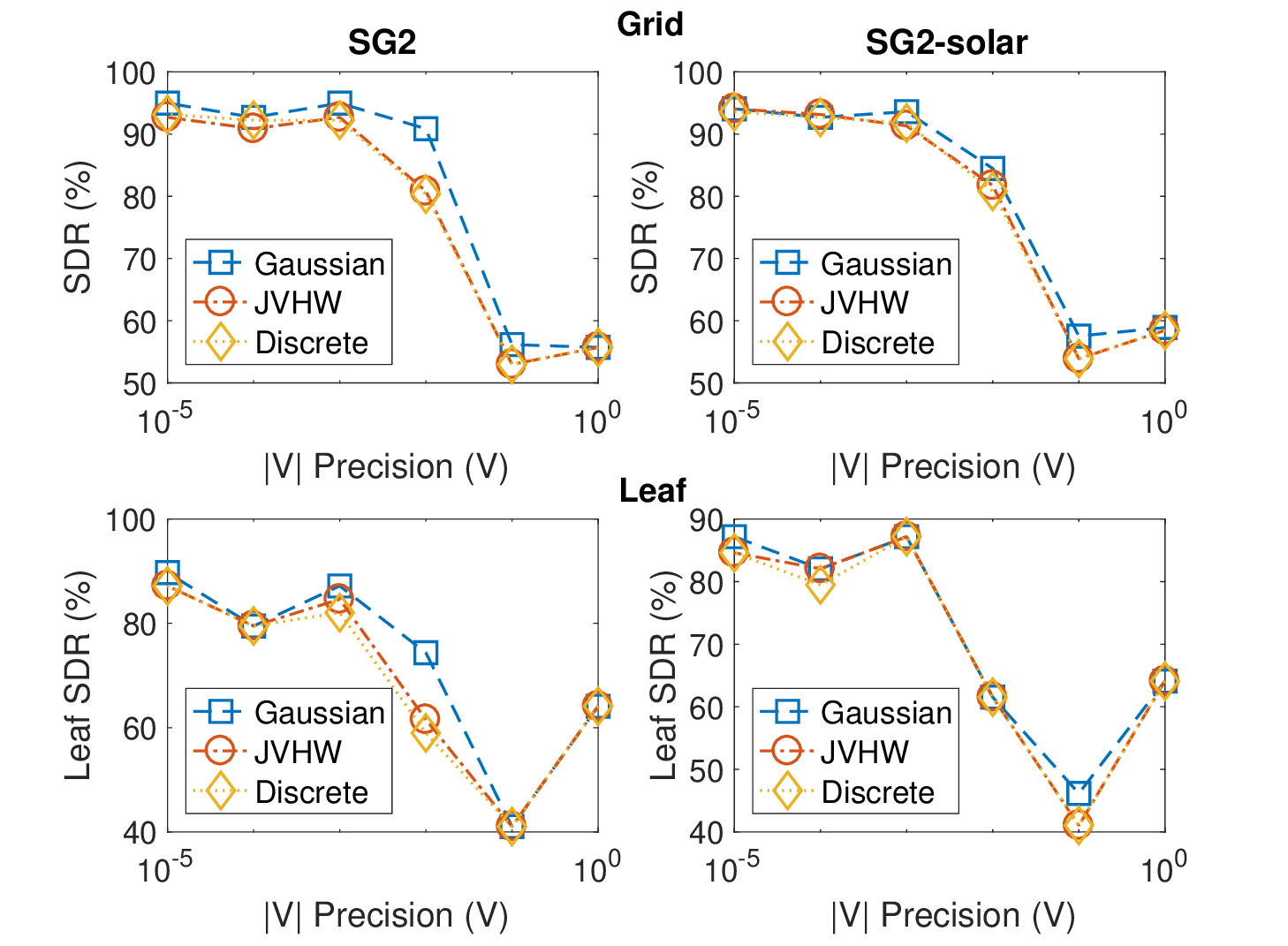}
\caption{The SG2 datasets are rounded to the nearest significant digit before analysis.}
\label{fg:SG2_sig_digs}
\end{figure}

In Figure \ref{fg:SG2_sig_digs}, we see for SG2 (non-solar), the Gaussian methods outperform other MI methods at all precision levels except 0 V. For SG2 (solar), the Gaussian method performs similarly to other methods and slightly better when less precise data is used. The leaf SDR is affected significantly by the number of significant digits. For SG2, the Gaussian performs the best. For SG2 (solar), the Gaussian method performs similarly to the other methods, except it outperforms the other methods when using precision of 10 micro-Volts of data and 100's of millivolts of data. We might expect the discrete MI methods to outperform the Gaussian method since the Gaussian approximation can be a crude approximation of the data. Both the SG2 and SG1 solar and non-solar results suggest precision of at least millivolts is required for estimation resulting in above $90\%$ SDR. 

\section{Discussion}

This work performs experiments on different datasets, different lengths of data, different down-sampling schemes, different levels of precision of data, and different methods to calculate the mutual information to visualize and quantify the performance of the Chow-Liu algorithm. The algorithm predicts the grid 100\% correctly for most IEEE datasets and for the SG1 datasets (GridLAB-D). The algorithm however finds more difficulty with the larger 200+ node network in SG2 datasets (GridLAB-D) with greater than 90\% of branches in the true system correctly estimated. The drop in performance for larger networks suggests the need for more experiments with SG2-datasets and other similar size datasets.

The SG1 and IEEE datasets connect leaf nodes correctly 100\% of the time based on mutual information in many settings. The SG2 datasets show worse performance with leaf nodes correctly connected greater than 85\% of the time. This drop in performance for SG2 datasets also suggests the need for more experiments using larger node networks. The reduced accuracy in identifying leaf nodes within larger networks can be attributed to the physical characteristics of voltage drop and error propagation~\cite{marulli2021reconstruction}. In extensive networks, the incremental change in voltage at leaf nodes is heavily attenuated by the cumulative impedance of long feeder lines~\cite{park2018exact}. Consequently, the mutual information signal-to-noise ratio diminishes, making it difficult for the Chow-Liu algorithm to distinguish between a leaf node's true parent and other topologically distant, but electrically similar, outlying nodes.

The algorithm shows reduced performance for down-sampled GridLAB-D data. The GridLAB-D datasets show large drops in performance from using data with samples taken every minute to down-sampled data with samples every hour with performance drop greater than 30\% in some cases. The IEEE datasets on the other hand show perfect 100\% topology estimation with samples taken every hour. More analysis is required to determine why IEEE datasets perform much better than GridLAB-D datasets at lower sampling frequencies. 

The length of data also appears to be important in correctly identifying the topology. We often see greater than a 5\% increase in correctly identifying the grid topology when using one month of data compared to one week of data from GridLAB-D datasets with samples taken every minute. The algorithm does not show much improvement after a certain length of time, for instance using a year's worth of data compared to half a year's worth of data (sampled every minute) shows little effect on estimation performance. The SG2 datasets also show trouble predicting certain nodes even when increasing the length of data considerably. Further work is suggested to ascertain common properties about consistently incorrectly predicted nodes in the SG2 datasets. The datasets show decent Gaussian performance but poor JVHW and discrete performance. The poor performance suggests further work is required to delve into the question of whether discretization is causing poor performance in JVHW and discrete MI methods.

The quality of data also has a large impact on algorithm performance. Experiments to reduce the number of bits with which to store voltage magnitude data show a greater than 10\% increase in topology estimation when using 8-bit data precision compared to 4-bit precision. From 8-bit precision to 16-bit precision we don't see a large increase in performance.

The experiments also trim data to a specified significant digit. These experiments show that retaining millivolts of voltage magnitude data precision compared to 10's of millivolts of precision can increase performance from 5\% to 15\%. This insight shows the limits of the algorithm and suggest further experiments to understand required data precision for sufficient performance.

While this study utilizes an information-theoretic baseline to establish data quality thresholds, future work should compare these limits against other topology estimation methodologies. Alternative approaches, such as pair-wise distance algorithms based on graph distances~\cite{deka2017structure, deka2023learning}, which have been shown to outperform standard graphical models, and structured norm minimization techniques~\cite{anguluri2021grid}, may exhibit different sensitivities to extreme data quantization. Furthermore, state-of-the-art machine learning methods, including graph neural networks (GNNs)~\cite{wu2020comprehensive}, transformers~\cite{mialon2021graphit}, and neural architecture searches~\cite{zoph2016neural, speckhard2023neural}, have found success in power grid modeling~\cite{varbella2024powergraph, mabrouk2022distribution}.

While GNNs generally outperform classical methods in predictive accuracy, the data fidelity thresholds identified in this study carry broad physical implications. Specifically our findings that 8 bit or millivolt level precision is sufficient for structural inference establish an optimistic theoretical baseline. Since these thresholds were derived using simulated GridLAB-D and MATPOWER datasets, they represent a controlled lower bound under ideal conditions. Real world advanced metering infrastructure telemetry is subject to noise, missing data, and asynchronous reporting~\cite{bhattarai2019big}. Therefore while our results suggest extreme high precision instrumentation may not be strictly necessary, future work must validate these quantization thresholds against field data to determine how much additional precision is required to overcome real world noise. Future research should also investigate whether the minimal data requirements identified here hold true when training more complex architectures like GNNs. Additional future work should focus on error frequency analysis for larger networks like the SG2 datasets, developing mixed Gaussian mutual information algorithms, and creating improved discretization methods. Furthermore, our reliance on the Chow-Liu algorithm explicitly constrains the reconstructed topology to a radial tree structure, as the algorithm strictly prohibits cycles. While distribution grids are predominantly radial, urban networks and microgrids increasingly employ meshed configurations to improve reliability. To apply an information-theoretic approach to meshed networks, future research must move beyond maximum spanning trees, potentially employing generalized graphical lasso~\cite{friedman2008sparse} or mutual information thresholding techniques that permit cyclic graph structures~\cite{deka2017structure}.

\section*{Data availability}

Two versions of the code are made available in Matlab and in Python. The Python repository can be found at \url{https://github.com/speckhard/python_grid_est}. An 8 node IEEE example is shown in a Jupyter notebook file in the example folder as well as an example main file. The repository can be installed using pip:~\url{https://pypi.org/project/grid_top_est/}. The MATLAB version can be found on GitHub~\url{https://github.com/speckhard/matlab_grid_estimation}. GIF files of the reconstruction program are also included to help the reader better understand the how the Chow-Liu algorithm reconstructs the grid. 

The datasets generated and analyzed during the current study are available via the DOI:~\url{https://doi.org/10.5281/zenodo.18135206}.

\section*{Acknowledgments}
We thank Dr. Yang and Prof. Rajagopal for their valuable feedback on the manuscript. We thank Dr. Yizheng Liao for his feedback on the analysis and the manuscript. We thank Sandia National Labs for their feedback on the analysis.

\section*{Funding}
This work received partial funding from the Stanford University Department of Civil and Environmental Engineering and the International Max Planck Research School for Elementary Processes in Physical Chemistry. We thank the Stanford School of Engineering for providing computational resources for this study.

\section*{Author Contributions}
D.T.S. conceived the study, developed the methodology and software, performed the formal analysis, and wrote the original draft and revised manuscript.

\bibliographystyle{ACM-Reference-Format}
\bibliography{grid_estimation.bib}

\clearpage
\appendix
\section{Additional Plots and Data}

The additional plots shown here serve to better understand the model performance.

\begin{figure}
\centering\includegraphics[width=0.7\linewidth]{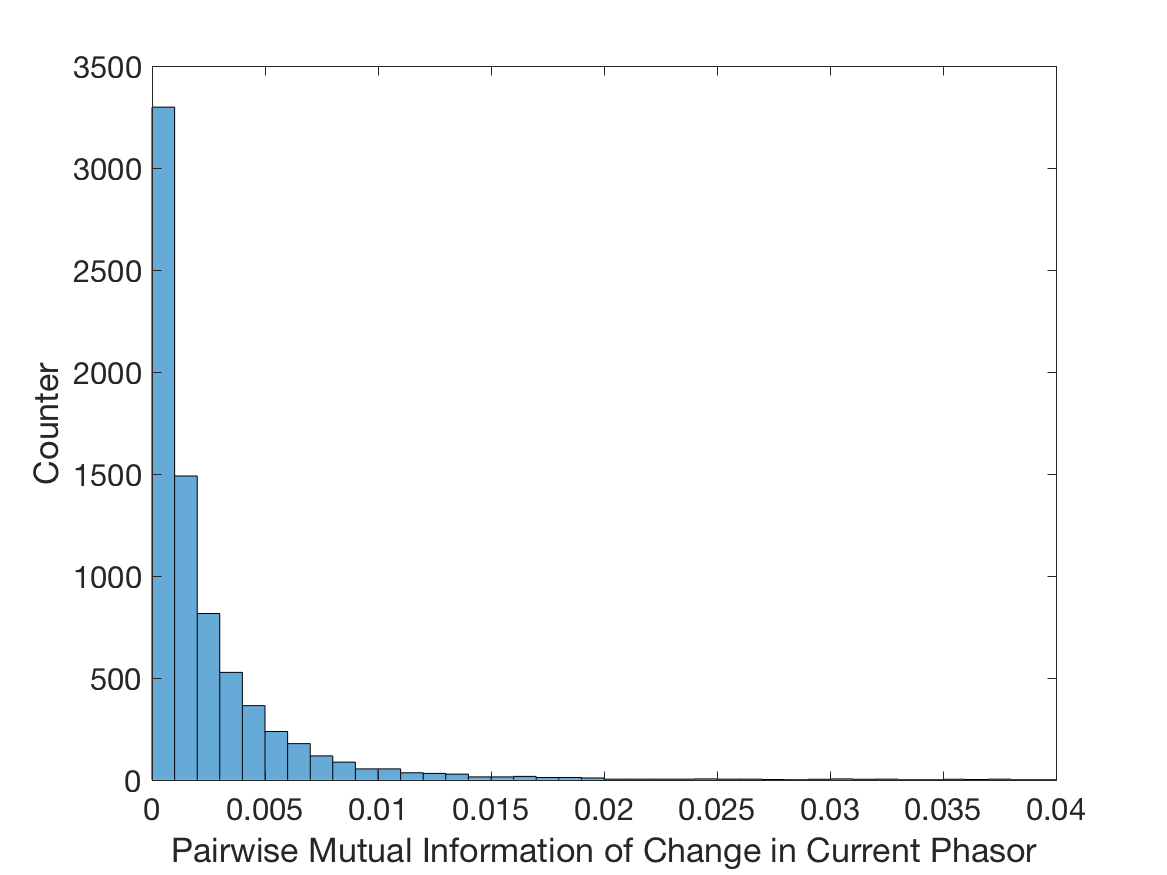}
\caption{The pairwise mutual information of current phasor for the SG1 (non-solar) dataset.}
\label{fg:current_MI}
\end{figure}

The pairwise mutual information of the change in the current phasor for the SG1 dataset is seen in Figure \ref{fg:current_MI}, The behavior appears to follow an exponential decay model.

\begin{figure}[h!]
\centering\includegraphics[width=0.8\linewidth]{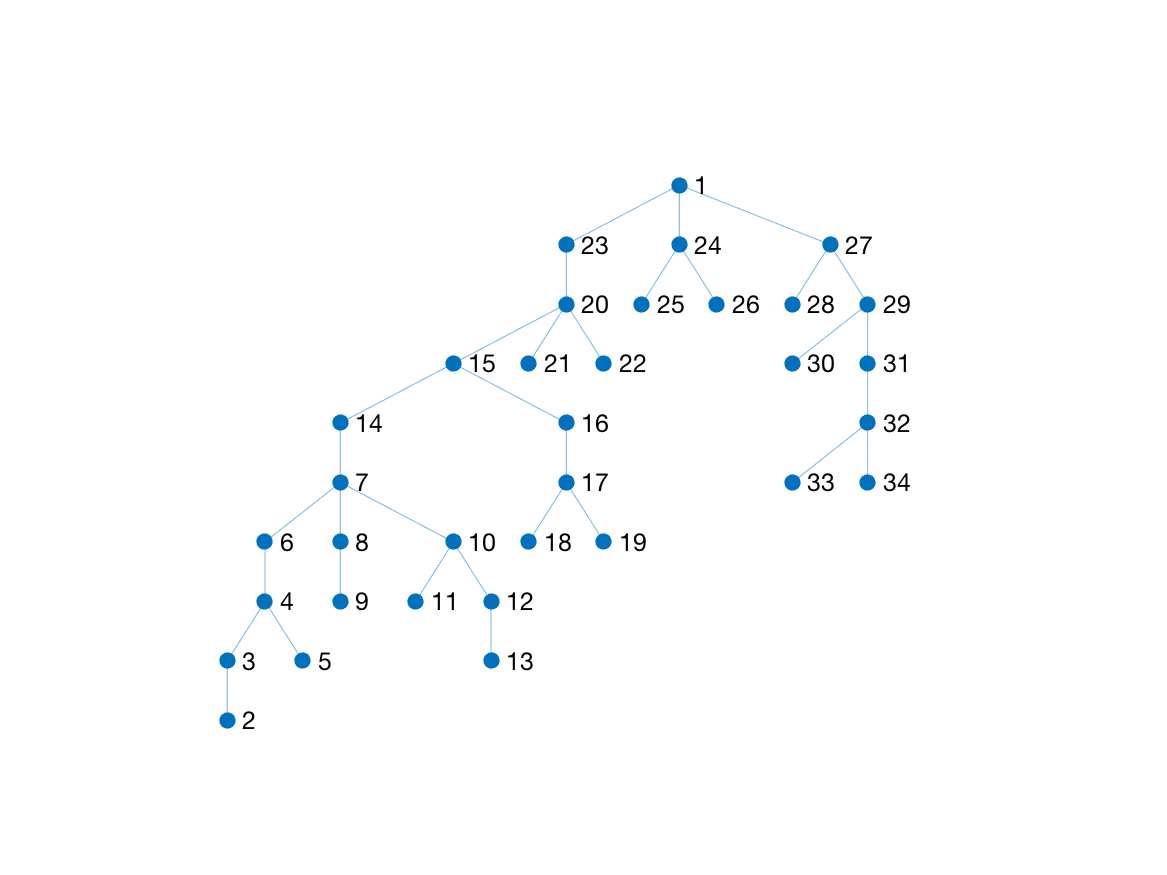}
\caption{The true graph of the SG1 dataset once redundant nodes are removed. This is the graph the algorithm will estimate and with which its performance will be compared.}
\label{fg:SG1_true_graph}
\end{figure}

\begin{figure}[h!]
\centering\includegraphics[width=0.8\linewidth]{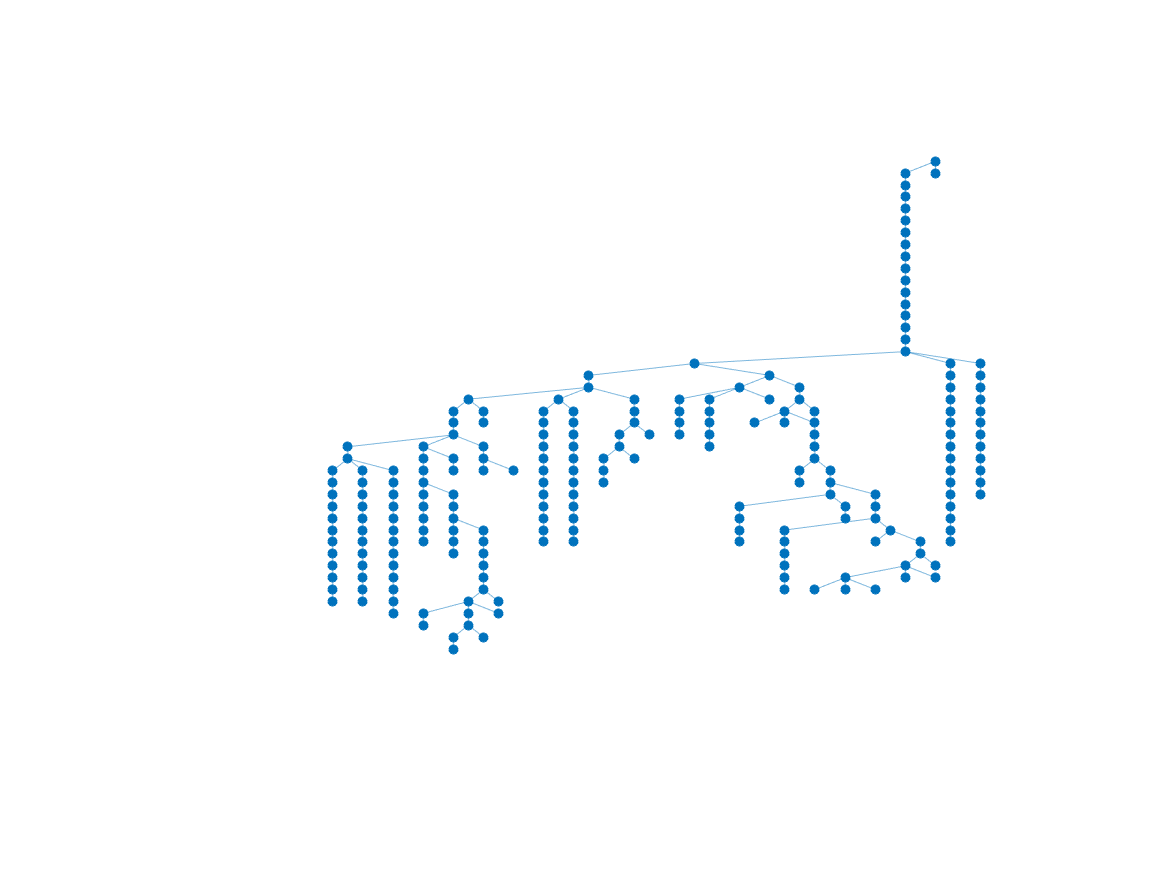}
\caption{The true graph of the SG2 dataset once redundant nodes are removed. This is the graph the algorithm will estimate and with which its performance will be compared. Note, this graph has a special layout and the text labels are removed since the graph has so many nodes that it is difficult to fit in labels and use a non-compact layout.}
\label{fg:SG2_true_graph}
\end{figure}

The SG1 dataset after the redundant nodes are removed is shown in Figure~\ref{fg:SG1_true_graph}. Similarly, the SG2 dataset after the redundant nodes are removed is shown in Figure~\ref{fg:SG2_true_graph}.

\begin{figure}[h!]
\centering\includegraphics[width=0.7\linewidth]{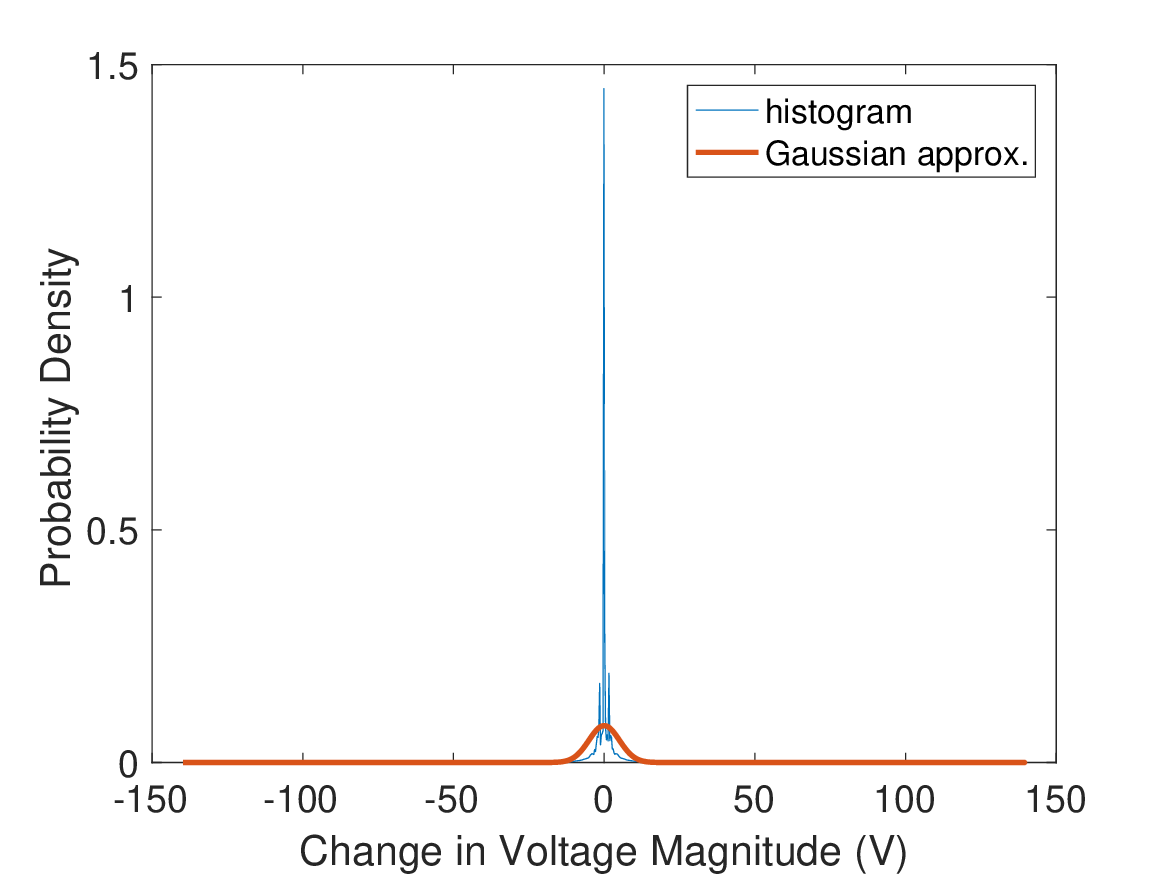}
\caption{The histogram and Gaussian approximation for the change in voltage magnitude at Node 33 in the SG1 non-solar dataset. A thousand bins are used for the histogram.}
\label{fg:sg1_node_33_hist}
\end{figure}

The histogram and the Gaussian approximation for the change in voltage magnitude from Node 33 in the SG1 non-solar dataset is shown in Figure~\ref{fg:sg1_node_33_hist}. Note, the figure looks slightly different than other histograms in this work since we use Matlab's histcount function so that we can more easily compare values from the histogram approximation of the function to those from the Gaussian approximation.

\begin{figure}[h!]
\centering\includegraphics[width=0.7\linewidth]{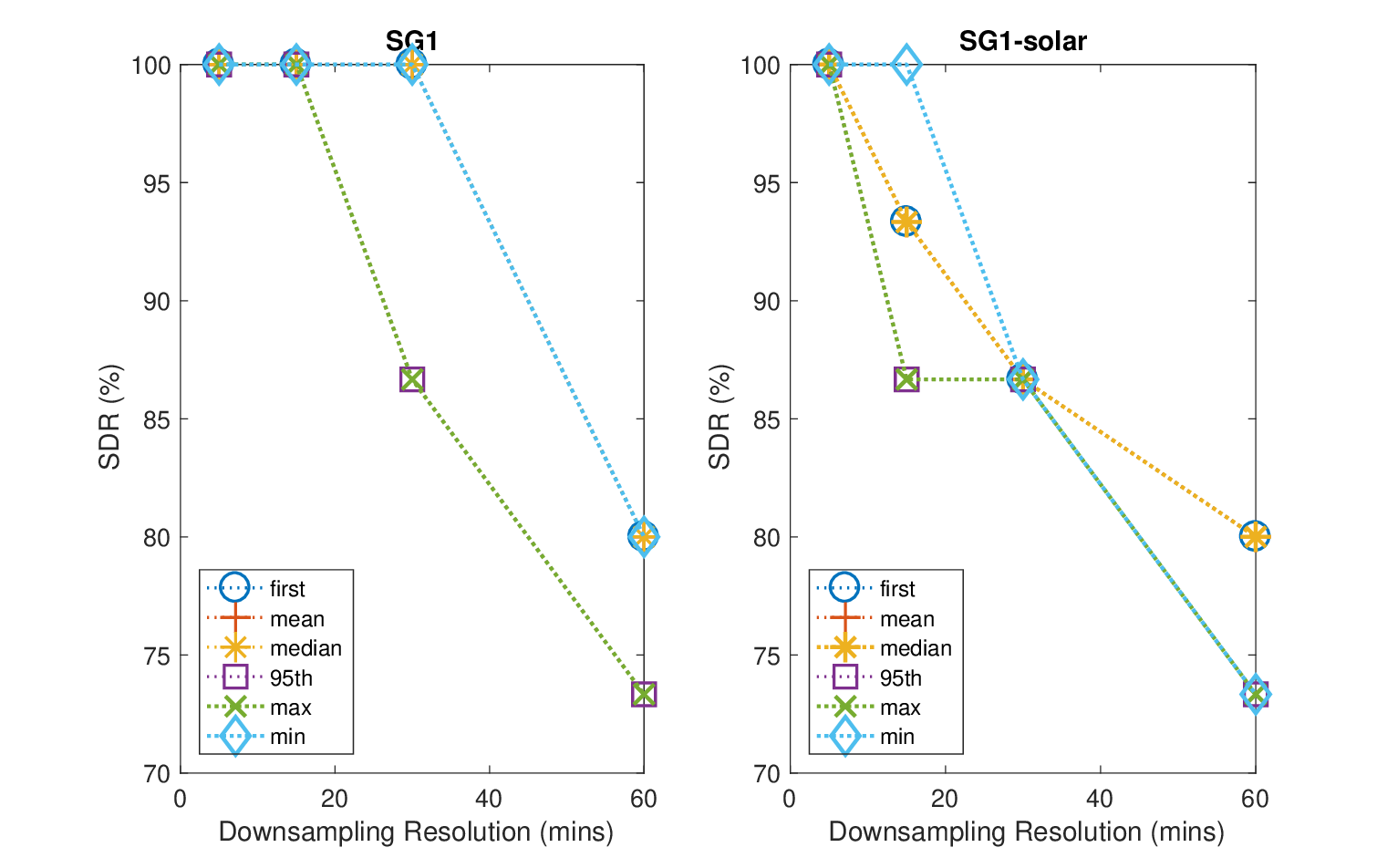}
\caption{The SG1 dataset is down-sampled from the original dataset containing one sample per minute to 1 sample per 5, 15, 30 and 60 minutes and then the change in voltage magnitude of the dataset is taken before estimation. Different down-sampling methods are compared on the full year dataset.}
\label{fg:SG1_dsample}
\end{figure}

The effect of different downsampling methods and downsampling frequencies on the SG1 dataset is shown in Figure~\ref{fg:SG1_dsample}.

\begin{figure}[h!]
\centering\includegraphics[width=1.0\linewidth]{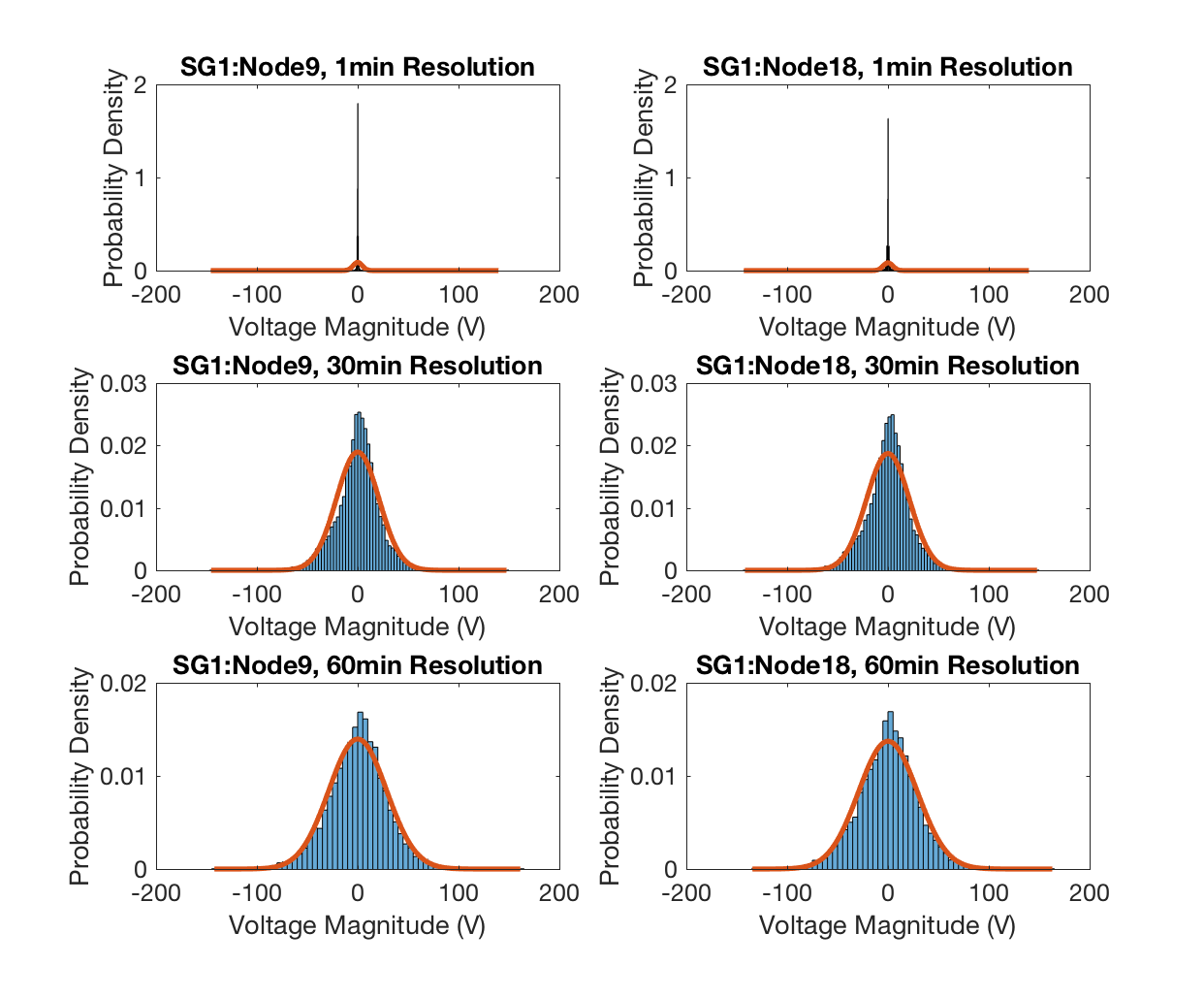}
\caption{Histograms and Gaussian fits (line in red) for the SG1 (non-solar) dataset for nodes 9 and 18 at different resolutions. The derivative has taken for all resolutions.}
\label{fg:SG1_hist_res}
\end{figure}

Histograms and Gaussian approximations for node number 9 and 18 at different time resolutions in the SG1 (non-solar) dataset are shown in Figure~\ref{fg:SG1_hist_res}.

\begin{figure}[h!]
\centering\includegraphics[width=1.0\linewidth]{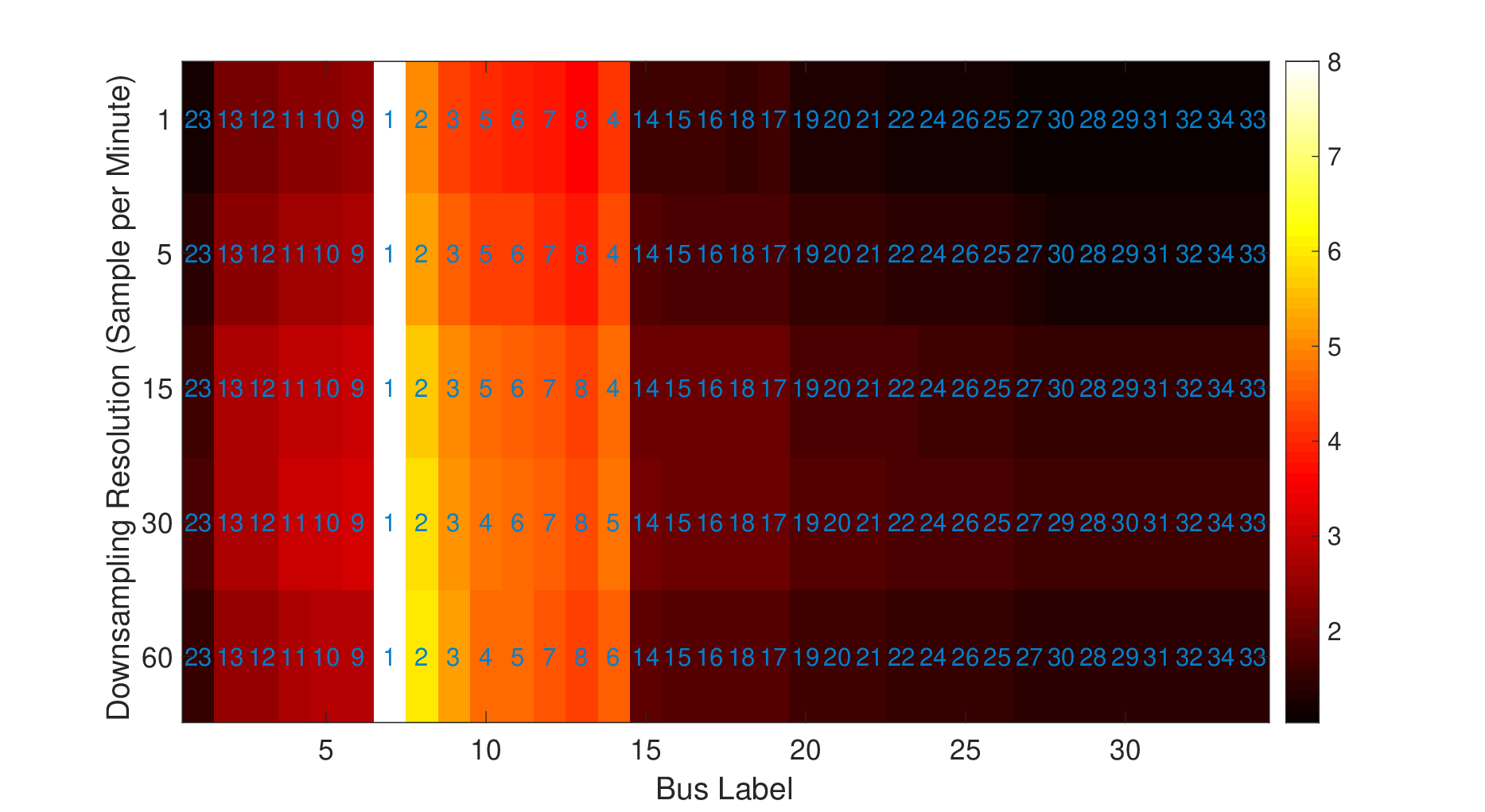}
\caption{ The mutual information pairing rankings as a function of downsampling resolution are shown as a heatmap for Node 7 in the SG1 dataset using the Gaussian mutual information on the change in voltage magnitude. The self-information, the mutual information between Node 7 and itself has the lowest ranking and conversely highest mutual information value for all resolutions. The resolutions shown are 1, 5, 15, 30 and 60 samples/minute.}
\label{fg:SG1_MI_dsample_heatmap}
\end{figure}

We can better understand the Chow-Liu algorithm and how it ranks pairwise mutual information values to connect nodes with edges when reconstructing the power grid graph by looking at Figure~\ref{fg:SG1_MI_dsample_heatmap}. The figure shows the mutual information pairwise rankings as a function of downsampling resolution for Node 7 in the SG1 dataset when calculating the mutual information using a Gaussian approximation.

\begin{figure}
\centering\includegraphics[width=1.0\linewidth]{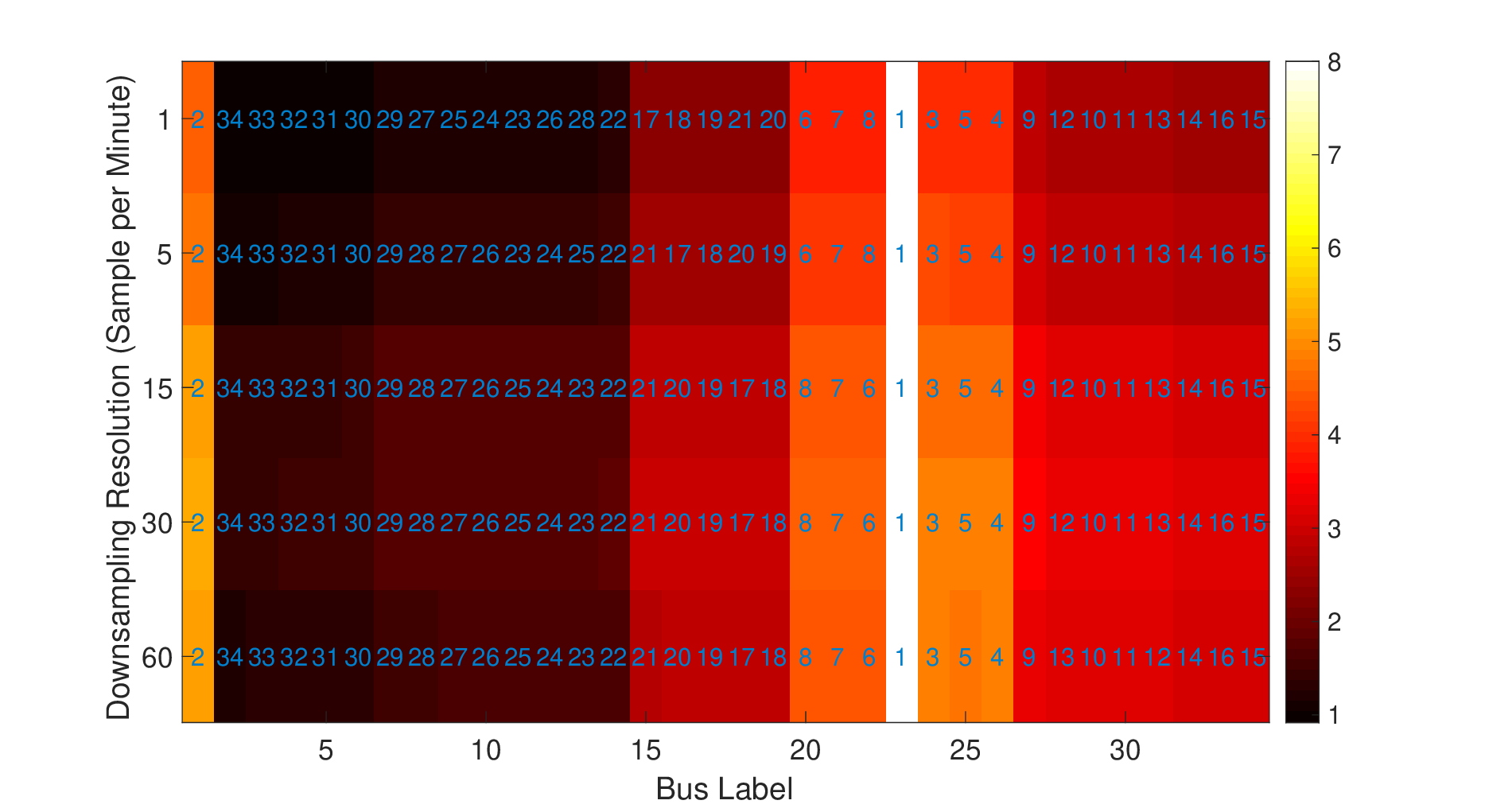}
\caption{ The mutual information paring rankings as a function of downsampling resolution are shown as a heatmap for Node 23 in the SG1 dataset using the Gaussian mutual information on the change in voltage magnitude. The self-information, the mutual information between Node 23 and itself has the lowest ranking and conversely highest mutual information value for all resolutions. The resolutions shown are 1, 5, 15, 30 and 60 samples/minute.}
\label{fg:SG1_MI_dsample_node23}
\end{figure}

Node 23's mutual information rankings in the SG1 dataset for the Gaussian approximations for different downsampling resolutions can be visualized in Figure~\ref{fg:SG1_MI_dsample_node23}. 

\begin{figure}[h!]
\centering\includegraphics[width=1.0\linewidth]{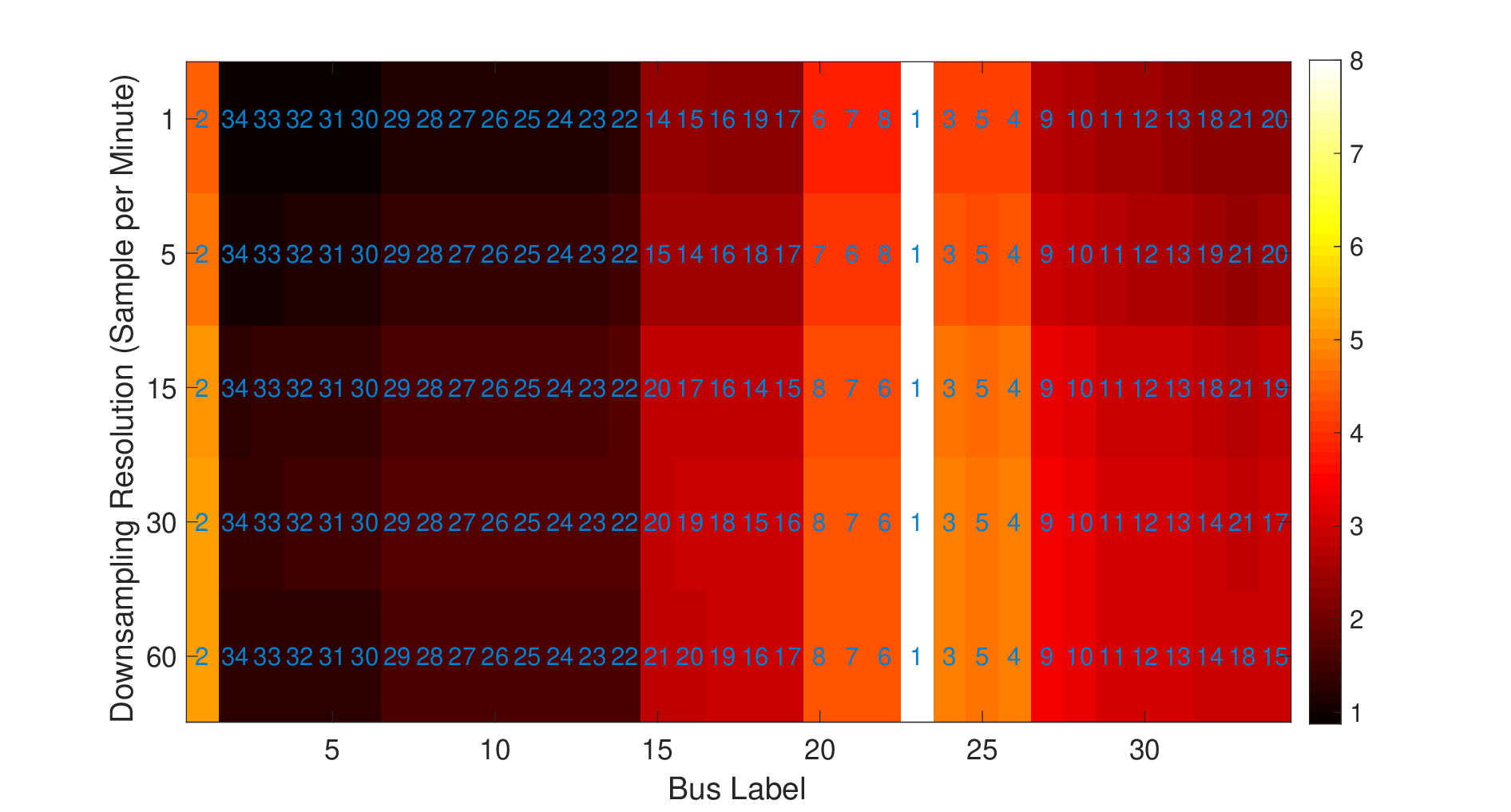}
\caption{ The mutual information paring rankings as a function of downsampling resolution are shown as a heatmap for Node 23 in the SG1 solar dataset using the Gaussian mutual information on the change in voltage magnitude. The self-information, the mutual information between Node 23 and itself has the lowest ranking and conversely highest mutual information value for all resolutions. The resolutions shown are 1, 5, 15, 30 and 60 samples/minute.}
\label{fg:SG1_solar_MI_dsample_node23}
\end{figure}

Figure~\ref{fg:SG1_solar_MI_dsample_node23} shows Node 23 for the SG1 solar dataset using the Gaussian approximation for different downsampling resolutions.

\begin{figure}[h!]
\centering\includegraphics[width=1.0\linewidth]{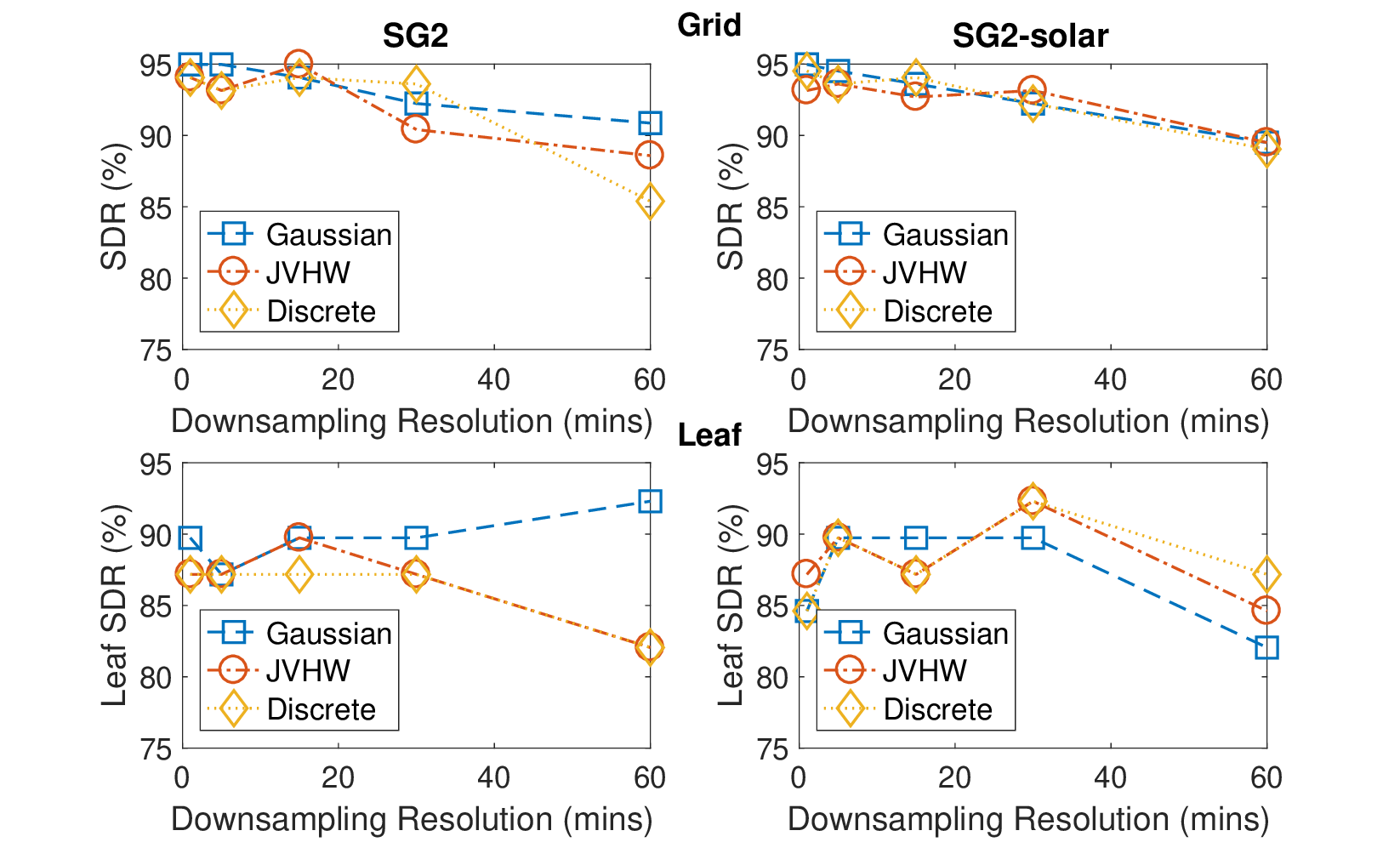}
\caption{The SG2 dataset is down-sampled from the original dataset containing one sample per minute to 1 sample per 5, 15, 30 and 60 minutes and then the derivative of the dataset is taken before estimation.}
\label{fg:SG2_res}
\end{figure}

The SDR of the SG2 dataset when varying the downsampling rate for different methods of calculating the mutual information is seen in~Figure~\ref{fg:SG2_res}.

\begin{figure}[h!]
\centering\includegraphics[width=0.7\linewidth]{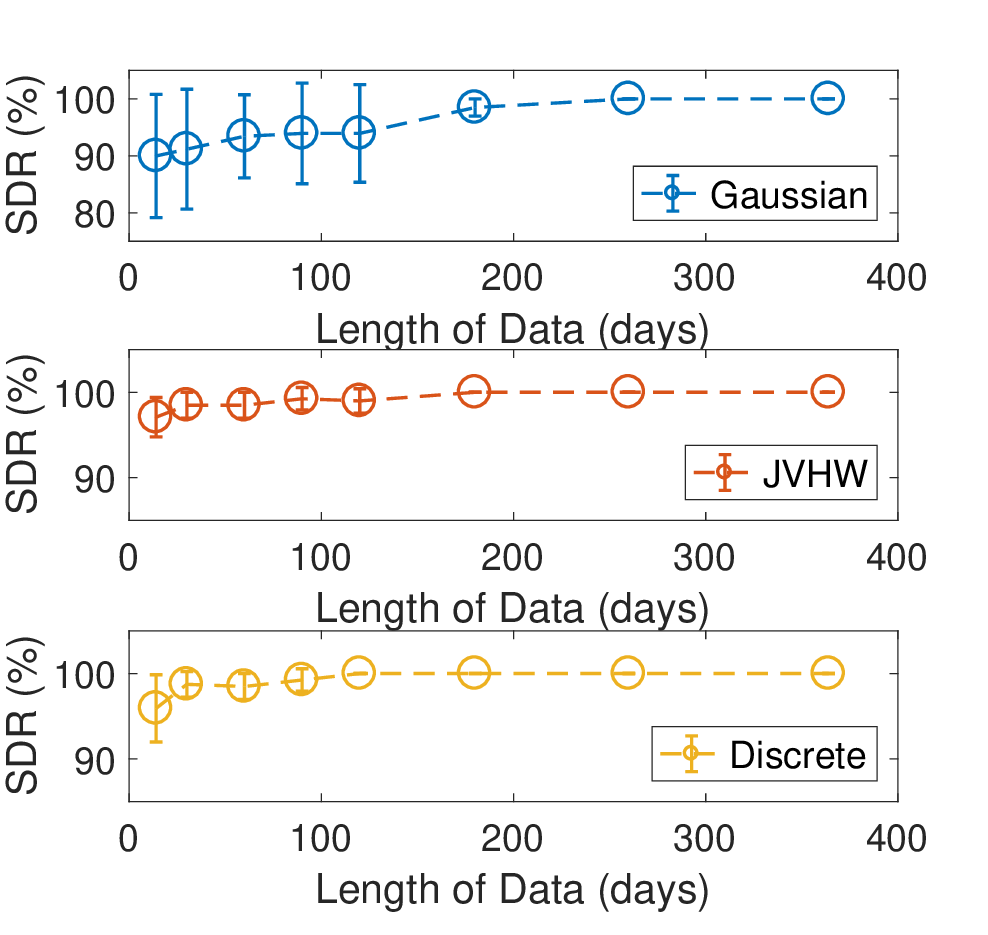}
\caption{Lengths of consecutive days of data are taken of the SG1 (non-solar) dataset before estimation.}
\label{fg:SG1_lens}
\end{figure}

The different time windows affect the SDR for the SG1 (non-solar) dataset are shown in Figure~\ref{fg:SG1_lens}.

\begin{figure}[h!]
\centering\includegraphics[width=0.7\linewidth]{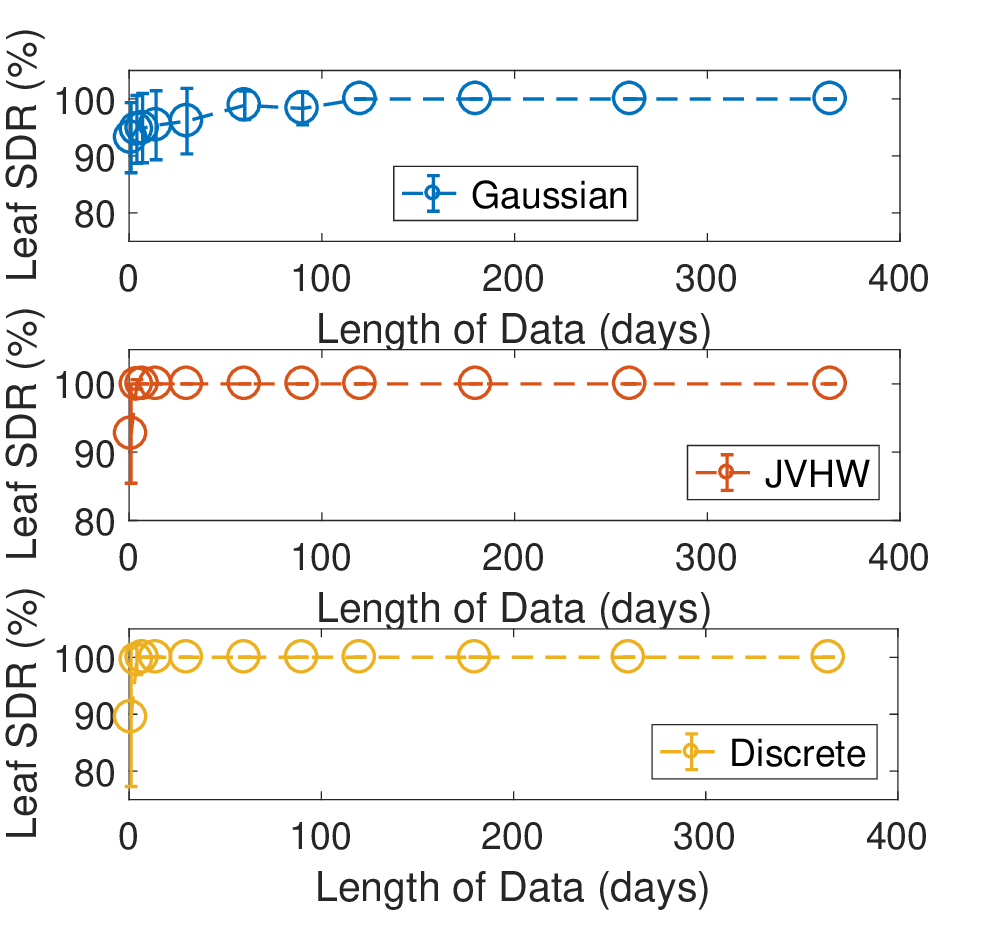}
\caption{Different lengths of consecutive days of data are taken of the SG1 (non-solar) dataset before estimation.}
\label{fg:SG1_lens_leaf}
\end{figure}

Figure~\ref{fg:SG1_lens_leaf} shows how the different time windows effect the leaf SDR for the SG1 (non-solar) dataset.

\begin{figure}[h!]
\centering\includegraphics[width=0.7\linewidth]{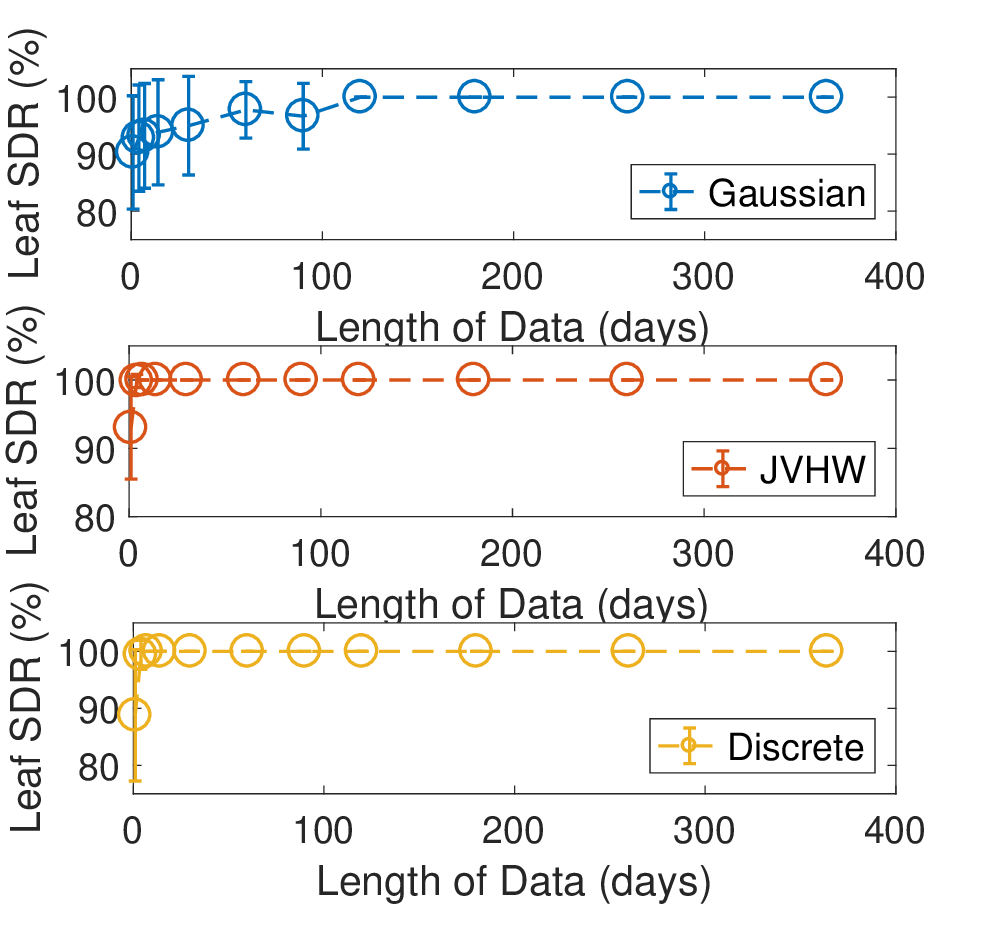}
\caption{Different lengths of consecutive days of data are taken of the SG1 (solar) dataset before estimation.}
\label{fg:SG1solar_lens_leaf}
\end{figure}

The effect of a variable time window length on the leaf SDR for the SG1 solar dataset is seen in Figure~\ref{fg:SG1solar_lens_leaf}.

\begin{figure}[h!]
\centering\includegraphics[width=0.7\linewidth]{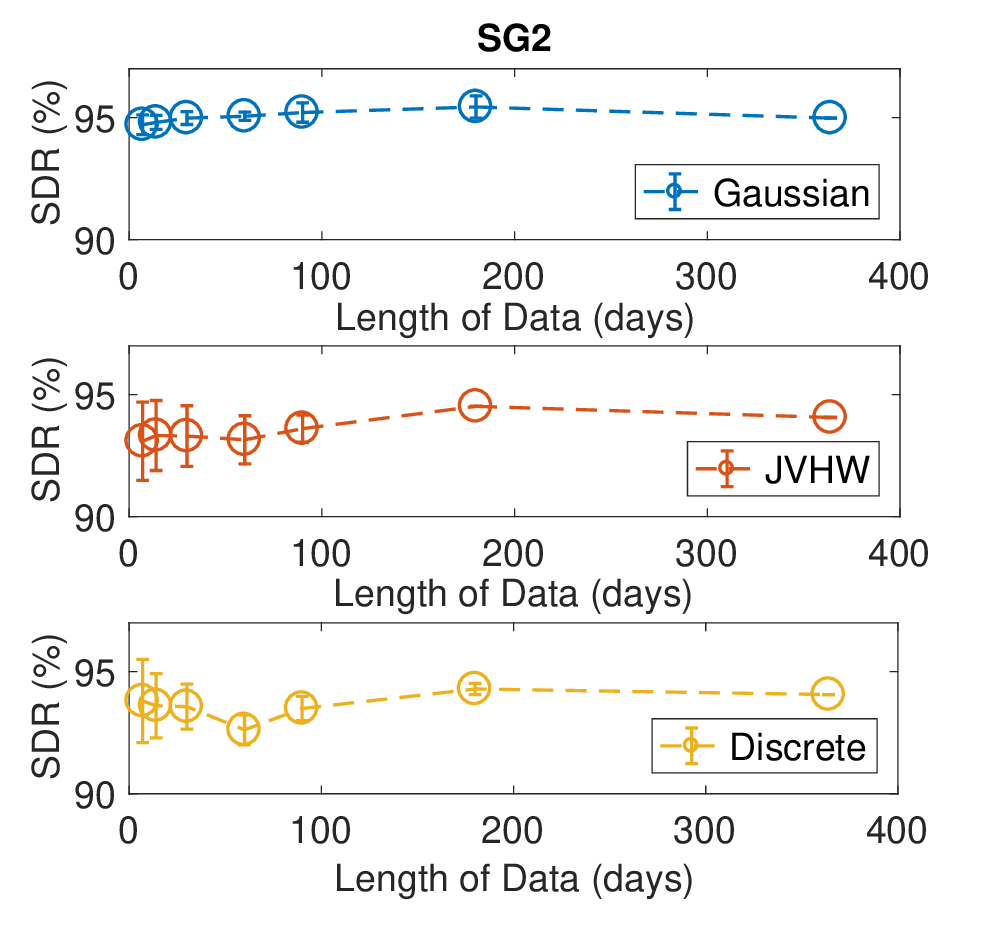}
\caption{Different lengths of consecutive days of data are taken of the SG2 (non-solar) dataset before estimation.}
\label{fg:SG2_lens}
\end{figure}

Figure~\ref{fg:SG2_lens} shows how the different time windows effect the SDR for the SG2 (non-solar) dataset.

\begin{figure}[h!]
\centering\includegraphics[width=0.7\linewidth]{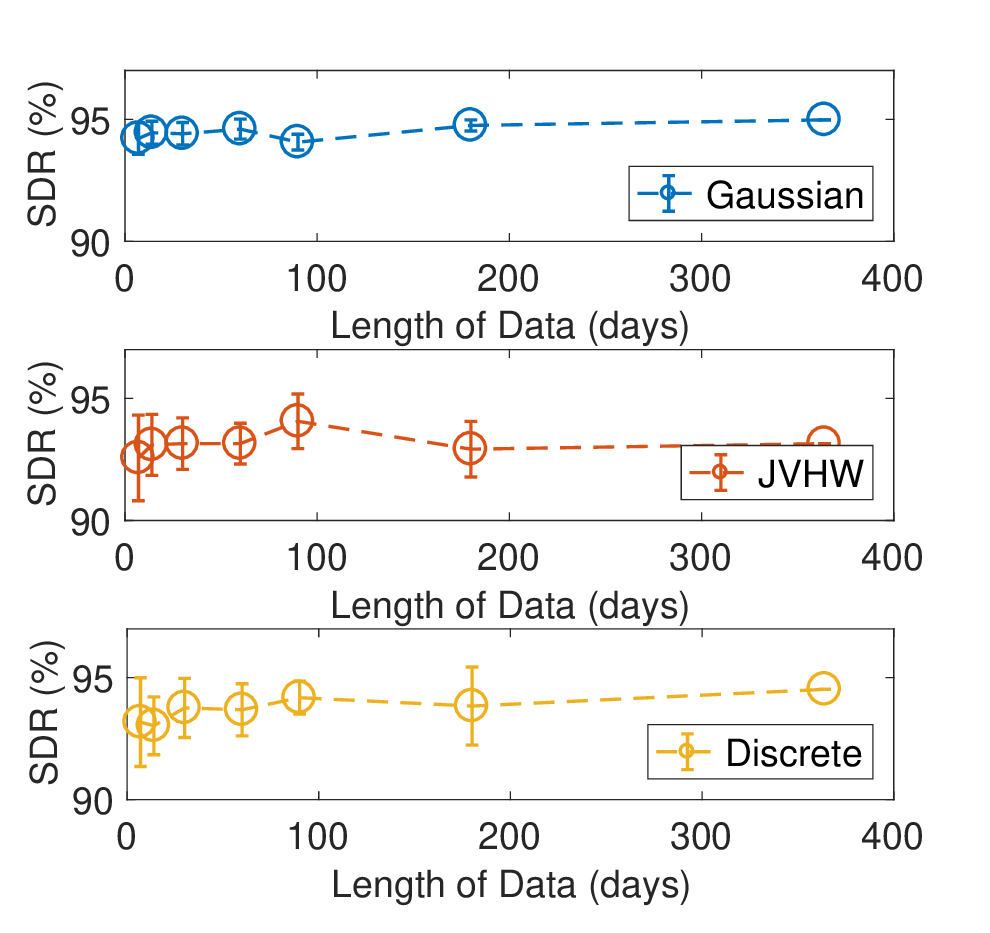}
\caption{Different lengths of consecutive days of data are taken of the SG2 (solar) dataset before estimation.}
\label{fg:SG2solar_lens}
\end{figure}

Different time windows plotted against the SDR for the SG2 solar dataset are shown in Figure~\ref{fg:SG2solar_lens}.

\begin{figure}[h!]
\centering\includegraphics[width=0.7\linewidth]{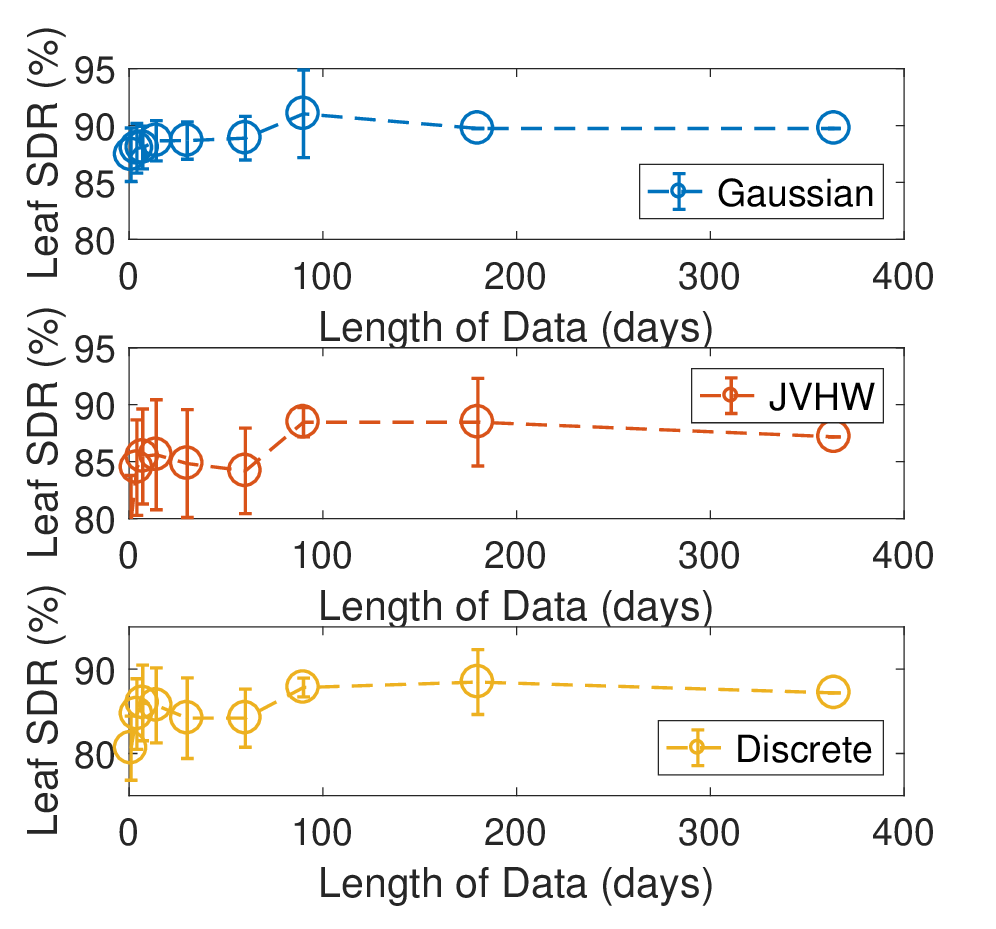}
\caption{Different lengths of consecutive days of data are taken of the SG2 (non-solar) dataset before estimation.}
\label{fg:SG2_lens_leaf}
\end{figure}

Figure~\ref{fg:SG2_lens_leaf} shows how the different time windows effect the leaf SDR for the SG2 (non-solar) dataset.

\begin{figure}[h!]
\centering\includegraphics[width=0.7\linewidth]{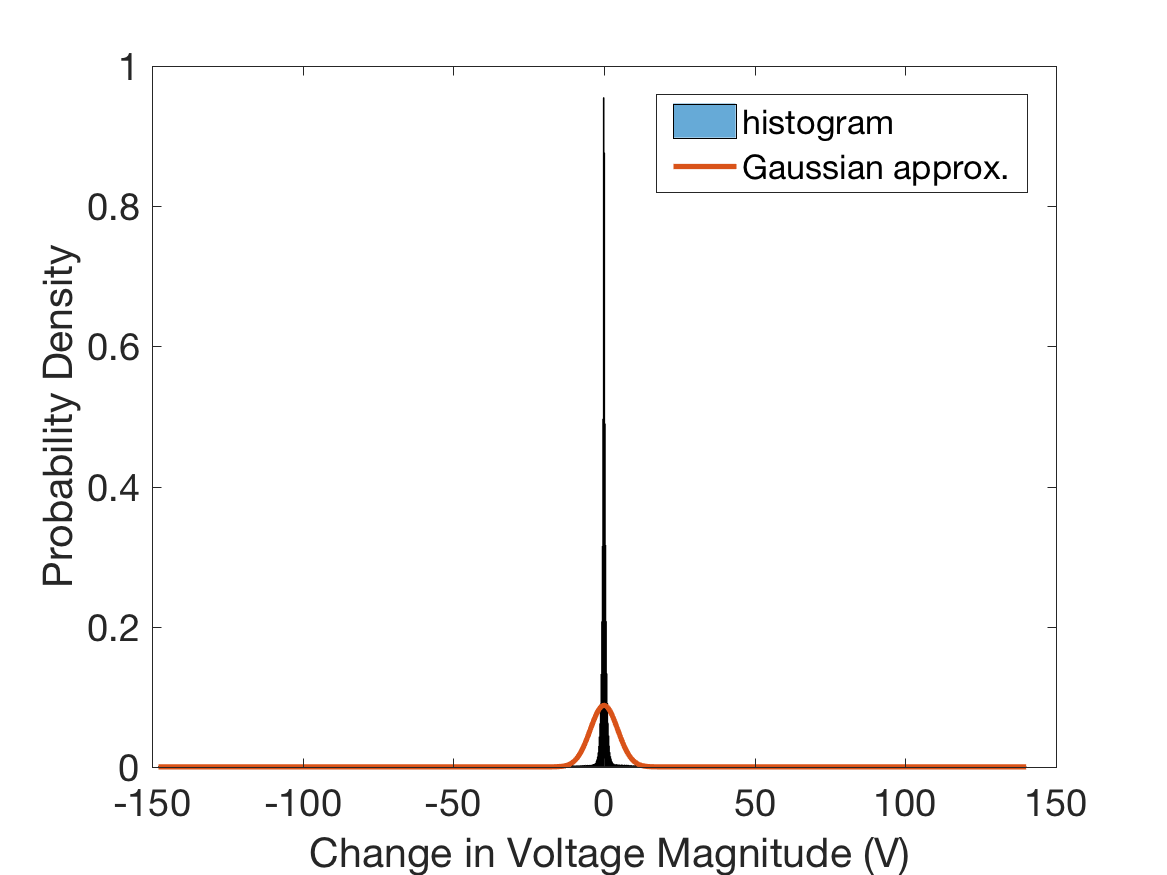}
\caption{Histogram and Gaussian approximation of the change in voltage magnitude data at Node 15 of SG2 (solar) dataset. }
\label{fg:SG2_solar_node15_deriv_hist}
\end{figure}

The histogram and Gaussian approximation of the change in voltage magnitude for Node 15 of the SG2 solar dataset is shown in Figure~\ref{fg:SG2_solar_node15_deriv_hist}. This node is consistently predicted to have an incorrect branch by the algorithm when using the Gaussian MI method for varying data lengths. This may be due to the Gaussian approximation being a poor fit as can be seen visually.

\begin{figure}[h!]
\centering\includegraphics[width=1.0\linewidth]{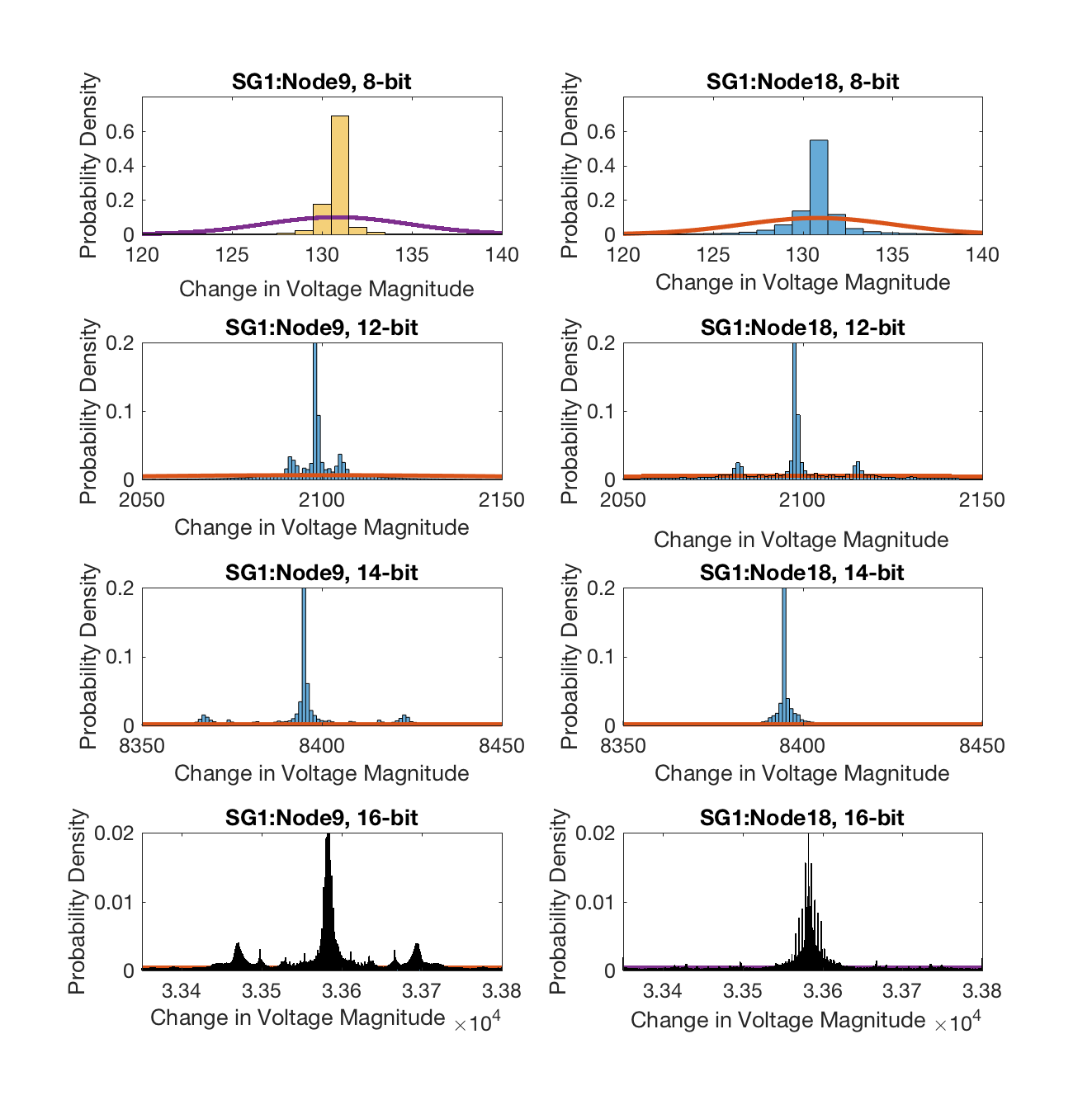}
\caption{The incremental change in voltage magnitude for 1 minute step sizes is taken for the SG1 dataset. The data is then discretized to a variable size in bits. Gaussian approximations are shown overlayed.}
\label{fg:SG1_hist_num_bits}
\end{figure}

Figure~\ref{fg:SG1_hist_num_bits} shows voltage magnitude data discretized for the SG1 dataset and their associated Gaussian approximations for Node 9 (left) and Node 18 (right).

\begin{figure}[h!]
\centering\includegraphics[width=0.7\linewidth]{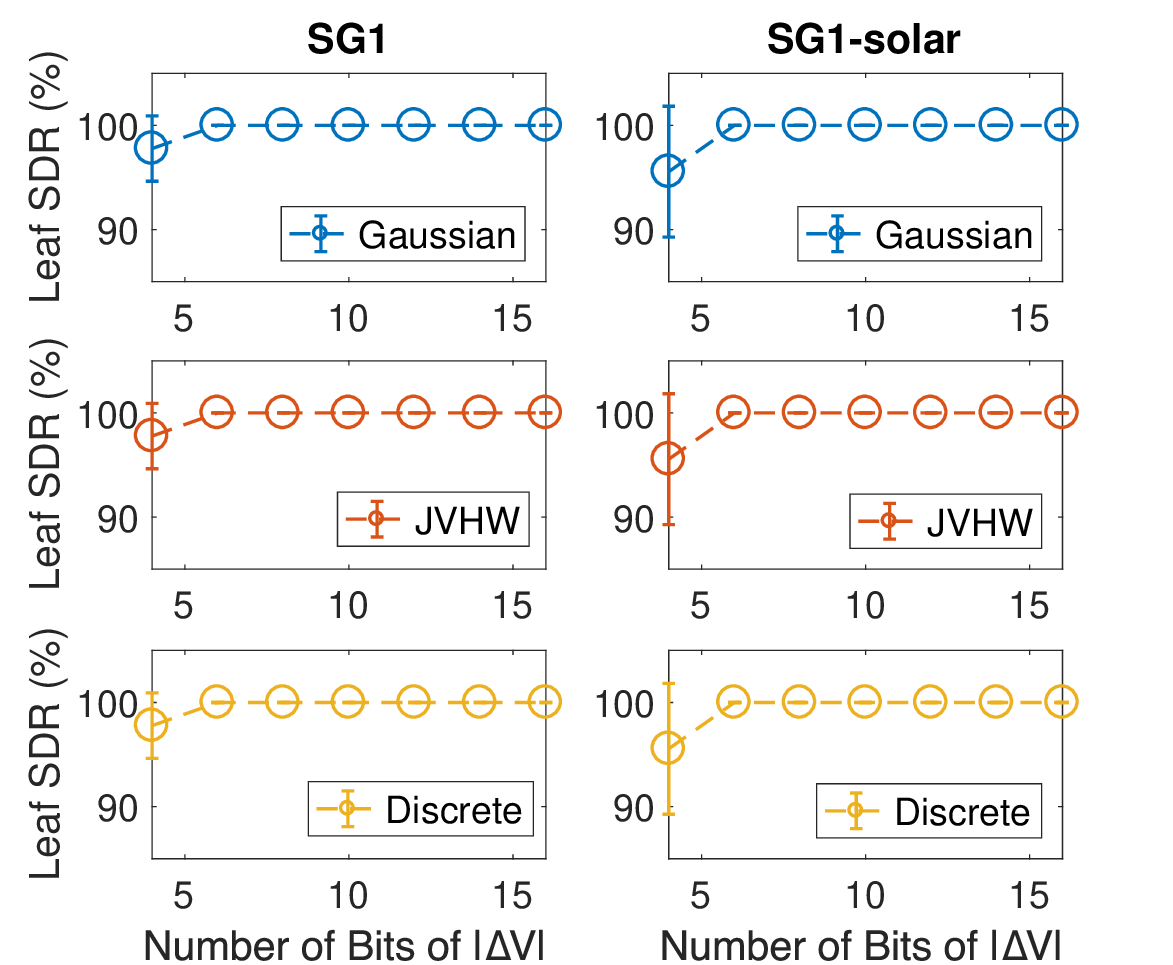}
\caption{The SG1 dataset is discretized to a certain amount of bits and the leaf SDR is analyzed. Here a length of 120 days is used. The lengths of data are taken consecutively in the year-long dataset, meaning three sets of 120 days are used in the analysis.}
\label{fg:SG1_num_bits_lens_leaf}
\end{figure}

The leaf SDR is analysis is shown for 120 day time windows for the SG1 dataset in Figure~\ref{fg:SG1_num_bits_lens_leaf}.

\begin{figure}[h!]
\centering\includegraphics[width=0.7\linewidth]{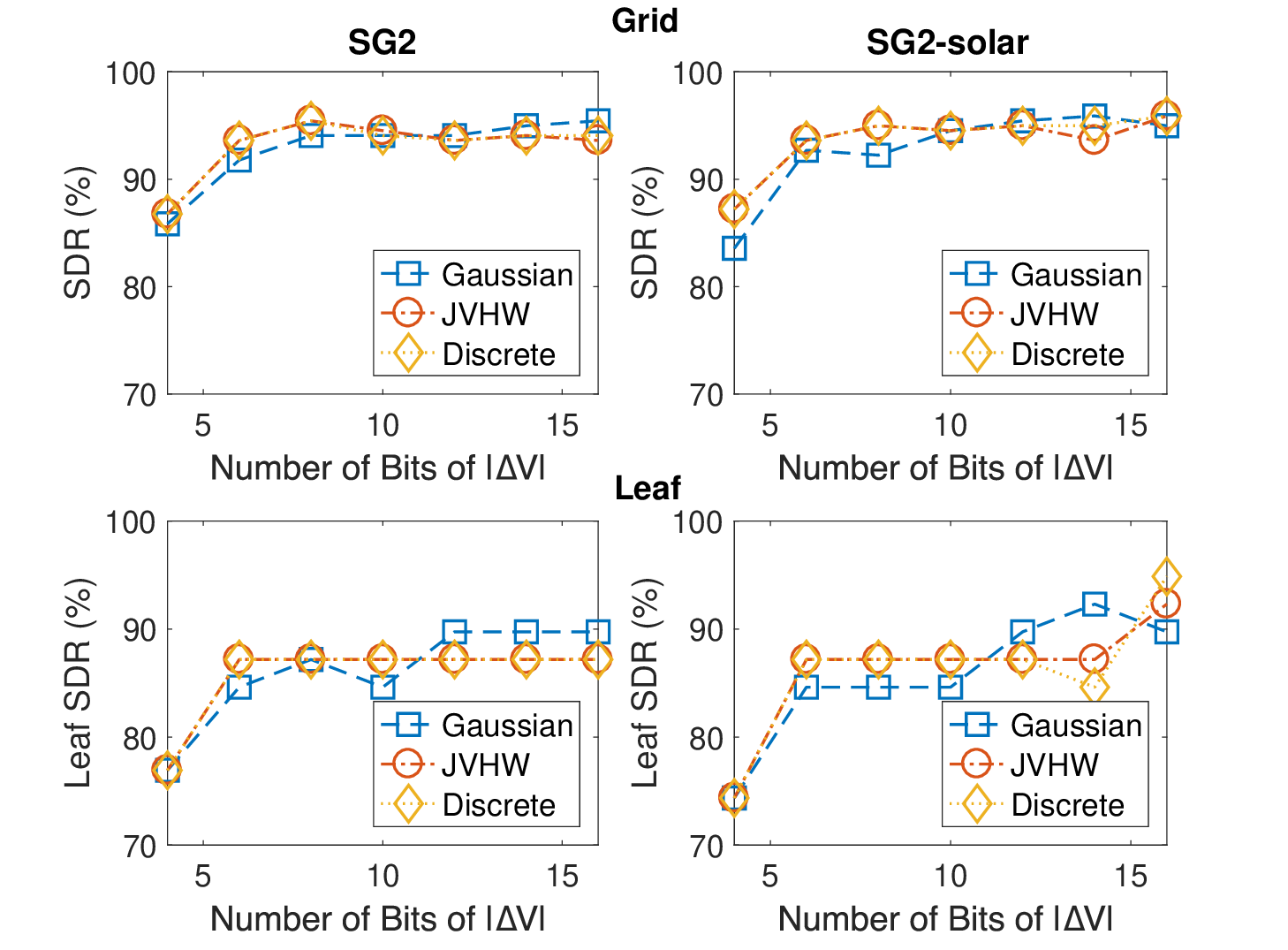}
\caption{The SG2 dataset is discretized to a certain amount of bits. The entire dataset is used for each MI method. }
\label{fg:SG2_num_bits}
\end{figure}

\begin{figure}[h!]
\centering\includegraphics[width=0.7\linewidth]{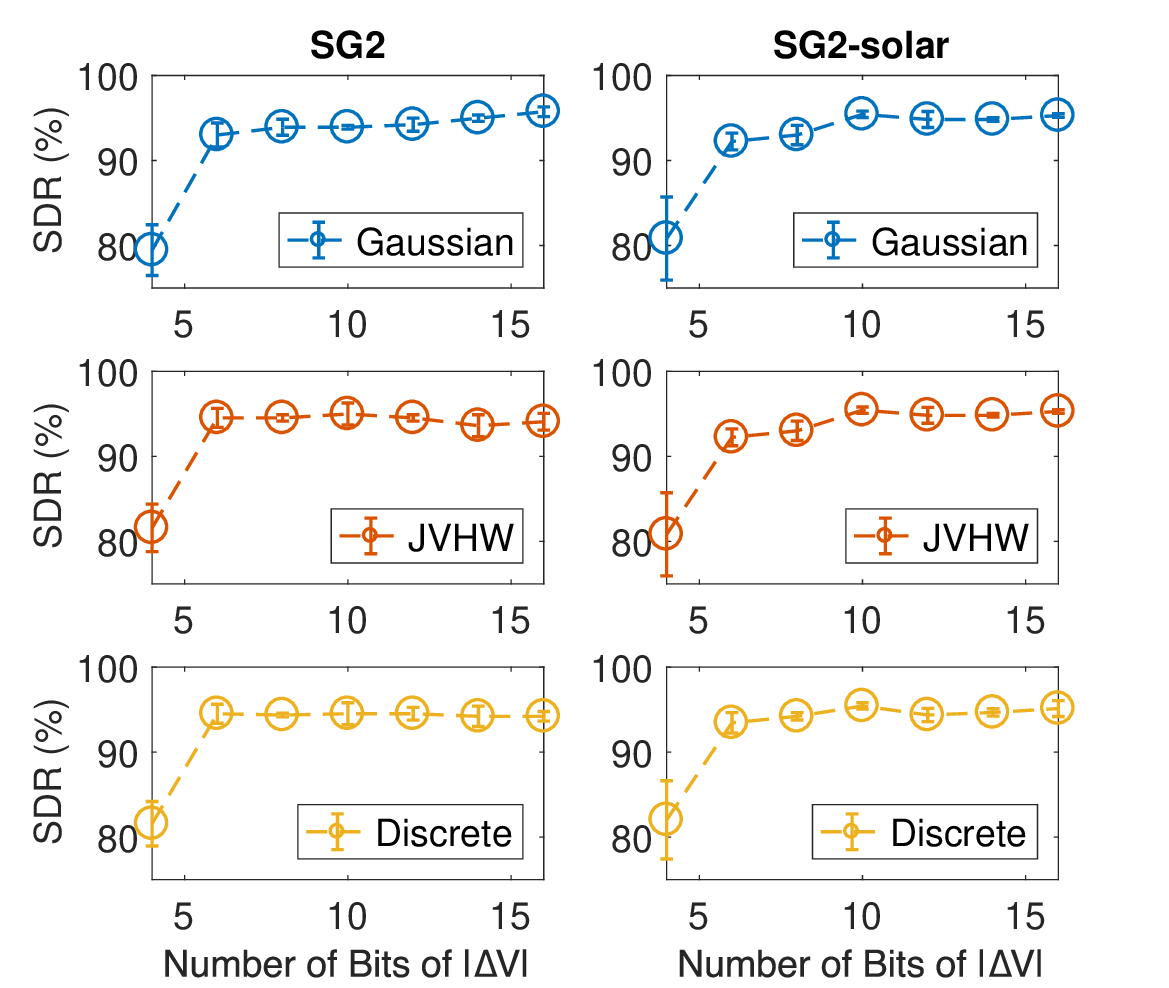}
\caption{The SG2 dataset is discretized to a certain amount of bits. Here a length of 120 days is used. The lengths of data are taken consecutively in the year-long dataset, meaning three sets of 120 days are used in the analysis.}
\label{fg:SG2_num_bits_lens}
\end{figure}

The SDR is analyzed is shown for the SG2 dataset in Figure~\ref{fg:SG2_num_bits}. For a 120 day window it is shown in Figure~\ref{fg:SG2_num_bits_lens}.

\subsection{Varying the Time Step-size of the Change of Voltage Magnitude Variable}

In this section we vary the time difference step size between voltage measurements and measure the effect on the algorithm's performance metrics. Equation (\ref{eq:step_size}) shows a step size of 2 minutes (where time is given in units of minutes) when calculating the change in voltage magnitude.
    
\begin{equation}
\label{eq:step_size}
\lvert \Delta V_{\text{Step Size 2}}[t] \rvert = \vert V[t] \rvert - \lvert V[t-2] \rvert 
\end{equation}

\begin{figure}[h!]
\centering\includegraphics[width=0.7\linewidth]{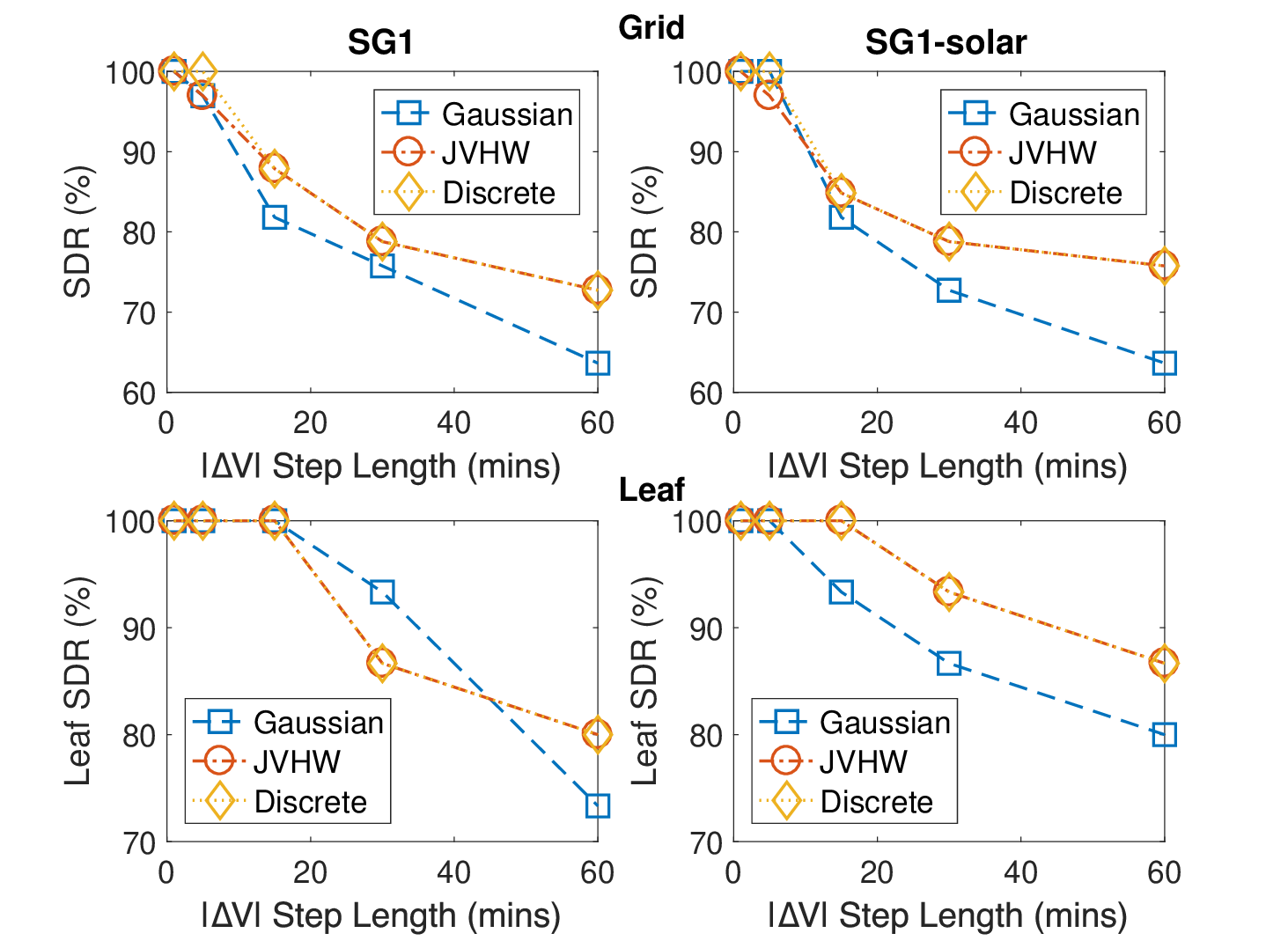}
\caption{Variable change in voltage magnitude step sizes of the SG1 datasets plotted against the successful detection rate (SDR) and leaf successful detection rate. Note, magnitude is abbreviated to mag. in the axis label.}
\label{fg:SG1_var_deriv}
\end{figure}

We can see in Figure \ref{fg:SG1_var_deriv} that both 34-node GridLAB-D datasets, SG1 and SG1 (solar), show significant drop offs in the SDR and leaf SDR for step sizes of 30 minutes or greater. The SDR drops off from around 100\% to less than 90\% for all MI methods for both SG1 datasets with 15 minute step-sizes. The discrete and JVHW methods estimate the SDR better than the Gaussian approximation as expected because the Gaussian approximation is a rather crude approximation for some nodes. The SDR plots show that change in voltage magnitude step sizes of greater than five minutes cause considerable degradation of estimation performance for the SG1 datasets. 

The leaf SDR plots show much better performance than the SDR plots in Figure \ref{fg:SG1_var_deriv} for all step sizes for the SG1 datasets. These datasets also show that fifteen minute step-sizes do not impact performance when using the discrete or JVHW MI methods. For the SG1 (non-solar) dataset, the Gaussian method outperforms the other MI methods by approximately $5\%$ at a step size of 30-minutes. The SG1 solar dataset shows the Gaussian method outperforms other MI methods in terms of leaf SDR for all step-sizes greater than 5 minutes. 

\begin{figure}[h!]
\centering\includegraphics[width=0.8\linewidth]{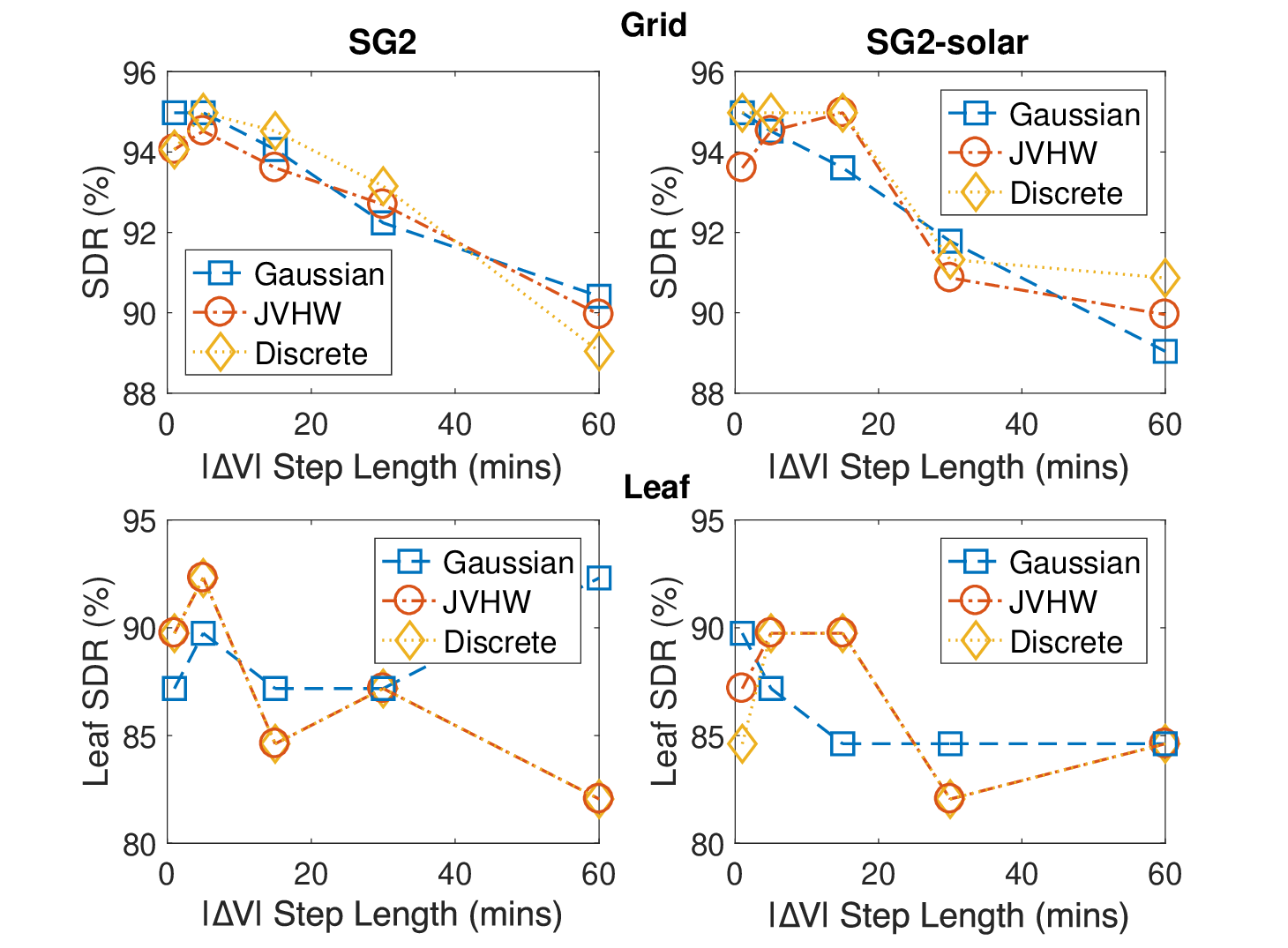}
\caption{Variable step sizes for the change in voltage magnitude for the SG2 datasets plotted against the successful detection rate (SDR) and leaf successful detection rate. Note, magnitude is abbreviated to mag. in the axis label.}
\label{fg:SG2_var_deriv}
\end{figure}

For the SG2 dataset, the Figure \ref{fg:SG2_var_deriv} shows the leaf SDR is lower than the regular SDR for variable changes in voltage magnitude step size. This is the opposite trend seen in the SG1 datasets. For the SG2 (non-solar) dataset the Gaussian MI method at a step size of 60 minutes performs much better than other methods in terms of leaf SDR. This is a trend seen in the resolution versus leaf node SDR experiments in the section below. In the SG2 (solar) dataset, the discrete and JVHW MI methods show improved SDR performance from a 1 minute step-length to a 5 minute step-length. The discrete and JVHW two MI methods also improve from a 30 minute step-length to a 60 minute step-length. 

\section{Kruskal's Algorithm}

Kruskal's algorithm seeks to find the minimum weight spanning tree. A tree is a graph with no cycles. In a tree you cannot draw a path using connections between nodes on the graph from one node and back to the same node. Each connection is referred to as a pair of nodes $(x,y)$ in the graph and has an associated numerical value known as the weight $w(x,y)$~\cite{cormen2022introduction}. The minimum weight spanning tree is the graph which connects the nodes so that the sum of the weights of the connections is minimized so long as there are no cycles in the graph.

Kruskal's algorithm sorts all of the node pair weights in increasing order. The nodes with the minimum weight are connected. The algorithm then sequentially connects the next smallest weightings so long as the connection does not create a cycle in the graph. Implementations of Kruskal's algorithm keeps track of the parent of each node. The parent or root of a node refers to the highest ranked node that is connected to the node in question.

There are several different implementations of Kruskal's algorithm, two of which differ in how they store the parent of a node~\cite{kershenbaum1972computing}. When a pair of nodes is connected in the algorithm, one node is assigned as the parent of the other. This means one node has a higher ranking than the other. A node with four parents could be referred to as a rank four node. Alternatively, if a node has more than one parent, the path to the lowest ranked node can be compressed where the algorithm only keeps track of the lowest ranked parent for each node. 

When adding a connection in the graph in Kruskal's algorithm we need to be sure the connection will not cause a cycle. If two nodes have the same parent node than a cycle will be formed if these nodes are connected. Thus, the algorithm, thus before adding a connection,s checks the lowest-ranked parent of each node about to be paired. The path compression implementation improves performance in this step of the algorithm. 

Matlab has its own min-span-tree function but it requires the bio-informatics toolbox which many labs do not have. In order to increase usage of the grid topology estimation algorithm we wrote our own script to avoid licensing issues. Our implementation performs 10x slower than Matlab's version as seen in Table \ref{tb:kruskal}. Since Kruskal's algorithm execution time is negligible compared to mutual information computation time (as will be soon shown), minimal effort has been placed to improve our implementation of Kruskal's algorithm.

\begin{table}[h!]
\centering
\renewcommand{\arraystretch}{1.4}
\captionsetup{skip=10pt}
\begin{tabular}{l c}
\toprule
\textbf{Implementation} & \textbf{Mean Performance (s)}\\
\midrule
Our Implementation & 0.0154 \\
Matlab & 0.0017 \\
\bottomrule
\end{tabular}
\caption{Timing performance for Matlab's and our implementation of Kruskal's algorithm on the random power factor IEEE-123 Node network. The mean of three runs was taken.}
\label{tb:kruskal}
\end{table}




\end{document}